\documentclass[11pt]{article}
\usepackage{amsmath,amssymb}
\usepackage{rotating}
\usepackage{caption}
\usepackage{young}
\usepackage[vcentermath]{youngtab}

\def\eo{\overset{_{\phantom{.}\circ}}{e}{}}

\def\go{\overset{_{\phantom{.}\circ}}{g}{}}

\def\Do{\overset{_{\phantom{.}\circ}}{D}{}}

\def\So{\overset{_{\phantom{.}\circ}}{S}{}}

\def\etao{\overset{_{\phantom{.}\circ}}{\eta}{}}
\def\zetao{\overset{_{\phantom{.}\circ}}{\zeta}{}}

\newcommand{\tM}{{\tt M}}
\newcommand{\tN}{{\tt N}}
\newcommand{\tP}{{\tt P}}
\newcommand{\tQ}{{\tt Q}}
\newcommand{\tR}{{\tt R}}
\newcommand{\tS}{{\tt S}}

\newcommand{\cV}{{\cal V}}
\newcommand{\cA}{{\cal A}}
\newcommand{\cB}{{\cal B}}
\newcommand{\cD}{{\cal D}}
\newcommand{\cQ}{{\cal Q}}
\newcommand{\cP}{{\cal P}}

\begin{document}

\begin{titlepage}

\hfill AEI-2013-226

\vspace{2.5cm}
\begin{center}

{{\LARGE  \bf Generalised geometry from the ground up }} \\

\vskip 1.5cm {Hadi Godazgar, Mahdi Godazgar
and Hermann Nicolai}
\\
{\vskip 0.8cm
Max-Planck-Institut f\"{u}r Gravitationsphysik, \\
Albert-Einstein-Institut,\\
Am M\"{u}hlenberg 1, D-14476 Potsdam, Germany}
{\vskip 0.4cm
\texttt{Hadi.Godazgar@aei.mpg.de, Mahdi.Godazgar@aei.mpg.de,\\
Hermann.Nicolai@aei.mpg.de}}
\end{center}

\vskip 0.35cm

\begin{center}
\today
\end{center}

\noindent

\vskip 1.2cm

\begin{abstract}
Extending previous work on generalised geometry, we explicitly construct an E$_{7(7)}$-valued vielbein in eleven dimensions that encompasses the scalar bosonic degrees of freedom of $D=11$ supergravity, by identifying new ``generalised vielbeine" in eleven dimensions
associated with the dual 6-form potential and the dual graviton. 
By maintaining full on-shell equivalence with the original theory at every step, our
construction altogether avoids the constraints usually encountered in other approaches to generalised geometry and, as a side product, also furnishes the non-linear ansatz for the dual (magnetic) 7-form flux for any non-trivial compactification of $D=11$ supergravity,
complementing the known non-linear ans\"atze for the metric and the 4-form flux.  
A preliminary analysis of the generalised vielbein postulate for the new vielbein 
components reveals tantalising hints of new structures beyond $D=11$ supergravity
and ordinary space-time covariance, and also points to the possible $D=11$ origins
of the embedding tensor. We discuss the extension of these results to E$_{8(8)}$.
\end{abstract}

\end{titlepage}

\section{Introduction}

Despite the fact that maximal $D=11$ supergravity \cite{CJS} has been known and much 
studied for more than three decades it is still not clear what the most efficient formulation
of the theory is, especially in view of the appearance of exceptional duality
symmetries under dimensional reduction and the relation of this theory
to the non-perturbative formulation of string theory, also known as M-theory.  Indeed, 
the recent discovery \cite{dWN13} of a new structure in $D=11$ supergravity, a new 
``generalised vielbein,'' is evidence of this; a development which was was triggered 
by the discovery of new SO(8) gauged supergravities in \cite{DIT}.
The new generalised vielbein was found 
in the context of the  SU$(8)$ invariant reformulation of $D=11$ supergravity 
proposed a long time ago \cite{dWNsu8}. In this reformulation the non-gravitational
degrees of freedom of the theory are used to extend the local (tangent space) symmetry by
replacing the local Lorentz group SO(1,10) by SO(1,3) $\times$ SU(8),
where the second factor coincides with the denominator of the duality coset 
E$_{7(7)}/$SU(8) that appears upon the reduction of $D=11$ supergravity 
to four dimensions \cite{cremmerjulia}. Similar ``generalised vielbeine'' had  
been found in a reformulation of $D=11$ supergravity 
appropriate for the reduction to three dimensions \cite{Nso16,KNS}. 

The SU$(8)$ reformulation is based on a $4+7$ split of eleven-dimensional space-time,  
where the fields are packaged in terms of objects that transform under local SU$(8)$ 
transformations in eleven dimensions by combining the gravitational and matter degrees
of freedom into single structures. Moreover, it is shown that the supersymmetry 
transformations of $D=11$ supergravity can be written in terms of these objects in a 
way that makes the local SO(1,3) $\times$ SU$(8)$ symmetry manifest. One particular 
SU$(8)$ covariant object in this reformulation is the generalised vielbein, which replaces the vielbein along the seven internal directions.  The new generalised vielbein found in \cite{dWN13} encompasses the 3-form along the 7-dimensional directions.   
A clear advantage of the SU$(8)$ reformulation is that it immediately yields the duality manifest Cremmer-Julia theory \cite{cremmerjulia} upon toroidal reduction. Moreover, it is also the appropriate framework in which to analyse the $S^7$ compactification of $D=11$ supergravity 
to maximally gauged supergravity in four dimensions \cite{dWNN8}. In fact, it is only within this framework that it has been possible to prove the consistency of the $S^7$ reduction \cite{dWNconsis, NP}, 
and to arrive at a workable formula for the full non-linear ansatz for the 3-form field 
(4-form flux) \cite{dWN13,GGN}. Indeed, one of the new results of the present work is 
that we now also obtain the non-linear ansatz for the dual 6-form field.

Somewhat independently of these earlier results, more recent attempts in viewing 
the fields of $D=10$ and $D=11$ supergravities in a unified way have centred on 
{\em generalised geometry}, again pointing to the importance of duality symmetries 
in the {\em unreduced} theory.  
Generalised geometry as originally proposed in \cite{Hitchin, Gualtieri} 
is based on an extension of the tangent space to include $p$-forms associated 
to the winding of branes sourcing $(p+1)$-forms, which ultimately allows for 
diffeomorphisms and gauge transformations to be combined in an enlarged 
symmetry group. In the most conservative applications of these ideas in the context of 
$D=11$ supergravity \cite{hullgenm, pachwald} the tangent space is enlarged to include 
windings of M2-branes, M5-branes and KK-branes. Meanwhile there are also proposals 
whereby the base space is also extended so that the fields depend not only on 
conventional coordinates, but also on winding coordinates \cite{westsl32, hillmann, BP}
(in fact, the association of new coordinates with central charges is an old idea).
A characteristic feature distinguishing these attempts from the earlier work of 
\cite{dWNsu8,Nso16,KNS} is that one usually has to impose restrictions on 
the coordinate dependence of the fields in order to realise the desired geometric
structures, whence the relation to the original $D=11$ supergravity becomes greatly obscured.

An early proposal for extending space-time, arising from the E$_{11}$ conjecture 
\cite{weste11}, is made in \cite{westsl32}, where it is suggested that there exists an 
extension of $D=11$ supergravity via a non-linear realisation of the semi-direct product 
of E$_{11}$ and its first fundamental representation $L(\Lambda_1).$ In this picture the 
fields are obtained from a level expansion of the E$_{11}$ algebra, while the coordinate 
dependence is controlled by $L(\Lambda_1)$ (thus in principle extending eleven-dimensional space-time to a space of infinitely many dimensions). Quite separately from the E$_{11}$ conjecture, the non-linear realisation method \cite{BO, west2000, locale11} gives 
a prescription for determining explicitly a given duality coset element in a particular 
representation. In order to test the E$_{11}$ proposal in the context of its finite-dimensional
E$_{7(7)}$ subgroup, and motivated by \cite{dWNcc}, this method was applied by 
Hillmann \cite{hillmann} to advocate an ``exceptional geometry''~\footnote{In fact, 
   this term was already used in \cite{KNS}.}
for $D=11$ supergravity. In this picture one considers a $4+56$-dimensional geometry 
where the dynamics of the fields in the 56-dimensional part is given by an 
E$_{7(7)}/$SU$(8)$ coset element. When these fields only depend on seven internal coordinates,
and the dependence on the space-time coordinates is dropped, this dynamics reproduces the dynamics of the fields in $D=11$ supergravity with components along the seven-dimensional directions assuming a duality relation between the 4-form field strength and 7-form field strength. Moreover, the supersymmetry transformation of the coset element is postulated to give rise to the fermionic degrees of freedom, in particular the gravitino along the 7-dimensional directions, which reproduces the supersymmetry transformations in the 7-dimensional part \emph{\`a la} \cite{dWNsu8} if dependence of the fields is again restricted to seven internal 
coordinates.  As we will argue, however, focusing attention only on the ``internal" part of the 
geometry, as is also done in other approaches to generalised geometry, may be 
too restrictive as this assumption is not even respected for the simplest non-trivial compactifications, as we will illustrate in Appendix C of this paper.

In the approach of \cite{BP}, the $D=11$ theory is viewed in a $7+4$ split of space-time,
with the four dimensions considered as ``internal".
In particular a sector of the theory is considered that contains fields along the 4-dimensional directions, which would correspond to internal directions in the usual way that the 
SL(5,$\mathbb{R}$) duality symmetry appears. Furthermore, the fields are taken to depend on the 4-dimensional directions and time. This sector of the theory is then formulated in terms of 
an SL(5,$\mathbb{R}$) ``generalised metric'' which arises from the membrane duality arguments of \cite{dufflu}. As in \cite{hillmann}, the formulation is based on a group theory element, but here the extension of the tangent space is considered in a generalised geometric language along the lines of \cite{Hitchin, Gualtieri, hullgenm, pachwald} and it is shown that the diffeomorphisms and 3-form gauge transformations in this sector of the theory are unified. However, unlike previous versions of generalised geometry, in Ref.\ \cite{BP} the 
base space is also extended so that the 4-dimensional 
space is enlarged to a 10-dimensional generalised space. This is to be viewed as the M-theory analogue of double field theory \cite{dft1, dft2, dftgenmet}. This analysis was extended to SO(5,5) in \cite{BGP} and to E$_{6(6)}$ and E$_{7(7)}$ in \cite{BGPW}, where the generalised metrics are found by a truncation of the E$_{11}$ non-linear realisation outlined above. The case of E$_{8(8)}$ is considered in \cite{GGPE8}, where an E$_{8(8)}$ matrix is found in 
terms of components of the vielbein, 3-form, 6-form and a new field in the 
same representation as the putative dual gravity field. However, the direct link to $D=11$ supergravity is lost, owing to the presence of this new field. 

A key aim of generalised geometry is to unify diffeomorphisms and gauge transformations 
in a single generalised diffeomorphism (as already proposed for E$_{8(8)}$ in \cite{KNS}).
Thereby, generalised geometry extends the notions of Lie derivative and bracket in a way that incorporates gauge transformations of form fields. If one considers an extension of the space-time coordinates as well, then the closure of the generalised transformations requires that the fields satisfy a duality covariant constraint, the section condition \cite{BGGP, CSW, BCKT}. This constraint can be viewed as a duality covariant restriction that allows one to reduce from the extended space to the usual number of dimensions. 
Furthermore, it has been shown \cite{CSW, CSW2} that generalised geometric formulations of eleven-dimensions can be realised in a more geometric setting akin to Riemannian geometry. Again considering a sector of $D=11$ supergravity, or equivalently considering space-times that are a warped product of Minkowski space and a $d$-dimensional manifold, 
in Ref.\ \cite{CSW} it is shown that the dynamics can be written in terms of a generalised Ricci scalar that is defined in terms of an associated generalised connection. Moreover, it is shown this structure also extends to the fermionic sector \cite{CSW2}.

While these approaches to generalised geometry  propose a radical reinterpretation, 
and, if successful, would  amount to a genuine extension of $D=11$ supergravity, 
we return here to the viewpoint \cite{dWNsu8} that it is the theory itself that points at directions 
in which progress can be made (a view supported by the fact that, in 35 years, 
no true field theory extension of $D=11$ supergravity has ever been found). 
Thus our approach remains grounded in $D=11$ supergravity such that at every stage of the construction the resulting structures remain {\em on-shell}
equivalent to the full $D=11$ supergravity both in the bosonic and the fermionic sectors
and such that at no point is any truncation or constraint on the coordinate dependence
of the fields required. 

In this paper, we demonstrate explicitly how a judicious analysis of the supersymmetry variations of fields in the SU$(8)$ reformulation of $D=11$ supergravity leads to new structures.  In particular, we find two other ``generalised vielbeine'' and show that together with the other two known from the literature, these generalised vielbeine are to be viewed as components of an E$_{7(7)}$ matrix in eleven dimensions.~\footnote{Such structures are already evident in Ref.\ \cite{KNS}, where it is shown that $D=11$ supergravity contains a $36 \times 248$ matrix that is part of a full E$_{8(8)}$ matrix in eleven dimensions.}  In addition, we embark on an understanding of these new structures and the consistency relations that they satisfy.  It is known \cite{dWNsu8} that the original generalised vielbein satisfies generalised vielbein postulates, which constrain its derivative along the four and seven-dimensional directions.  These consistency requirements are a crucial ingredient in understanding the relation 
between the maximal gauged supergravity in four dimensions and the $D=11$ theory and in proving the consistency of the reduction to the former \cite{dWNconsis,NP}.  We present similar generalised vielbein postulates satisfied by the other generalised vielbeine. Recall that in Riemannian geometry the vielbein postulate is the requirement that the vielbein be covariantly constant, which gives an equivalence between the affine and spin connections that are defined on two different bundles. The fact that the generalised vielbeine satisfy analogous relations with more general connections is strong indication of the emergence of structures beyond Riemannian geometry.  Furthermore, at a more practical level, we expect that a deeper study of these relations will lead to the exciting possibility of understanding the higher dimensional origins of the embedding tensor \cite{NSmaximal3, NScomgauge3, dWSTlag}, which is the most efficient way of understanding 
gauged supergravities in any dimension.  

Another bonus of these results is that they give a non-linear ansatz for the dual 6-form potential.  Indeed, by considering the Englert solution \cite{englert} as a simple example and showing that not only is the 6-form field non-zero in this case, but that it has non-vanishing mixed components, we argue that this will generically be the case for all compactifications with non-vanishing flux. We believe this is an important point that one must bear in mind in any study involving dual fields in the context of $D=11$ supergravity, or any truncation thereof.  The results presented here are relevant for the $4+7$ split of the eleven-dimensional theory corresponding to the E$_{7(7)}$ duality group. As emphasised before, our analysis is based on the fermionic sector, in contrast to the mainly bosonic approach in the generalised geometry literature. Furthermore, eleven dimensional dualisation of the fields plays an important role in this story.  Finally, we also outline how similar structures can be constructed for 
cases relevant for other duality groups, in particular for E$_{8(8)}$.


In section \ref{sec:duality}, after a brief review of $D=11$ supergravity and the supersymmetry transformations satisfied by its fields, we motivate the importance of dualisation of fields in eleven dimensions in any attempt to understand duality symmetries from a higher dimensional point of view.  In particular, we emphasise the significance of the supersymmetry transformations of dual fields in the context of this work.  We derive the supersymmetry transformation of the six-form potential dual to the three-form potential of $D=11$ supergravity in section \ref{sec:6formsusy}.  Furthermore, we highlight the problems associated with a consistent \emph{covariant} and \emph{Lorentz invariant} formulation of dual gravity in general, but also in the context 
of eleven dimensions in section \ref{sec:dualgrav}.

Working within the context of the SU$(8)$ invariant reformulation of Ref.\ \cite{dWNsu8}, in section \ref{sec:e711d} we construct an E$_{7(7)}$ matrix in eleven dimensions that encompasses the bosonic degrees of freedom of the eleven-dimensional theory.  In particular, in section \ref{sec:newgenv}, we demonstrate the existence of another generalised vielbein in addition to the two previously known in the literature \cite{dWNsu8, dWN13}.  We argue in section \ref{sec:gengeoe7} that these generalised vielbeine must form the components of a single E$_{7(7)}$ valued object in eleven dimensions---a 56-bein, and conclude that the missing component must be related to a dual gravity field.  We construct this final generalised vielbein by insisting that it too transform as an E$_{7(7)}$ object.  Furthermore,  we show that the vector fields whose supersymmetry transformations give the generalised vielbeine can themselves be combined into a {\bf 56}-plet of E$_{7(7)}$.

Section \ref{sec:gvp} is devoted to a preliminary analysis of the generalised vielbein postulates satisfied by the new generalised vielbeine given in Ref. \cite{dWN13} and in section \ref{sec:e711d}.  The new generalised vielbein postulates give rise to as yet unknown connections associated with $p$-form gauge transformations.
Finally, in section \ref{e8e6}, we briefly discuss how one can implement a similar construction for the $3+8$ split of eleven dimensions, which would be relevant for the E$_{8(8)}$ duality group and also for the $5+6$ split relevant for the E$_{6(6)}$ duality group.  

%
%

\section{$D=11$ supergravity and duality} \label{sec:duality}

The lagrangian of eleven-dimensional supergravity \cite{CJS} in the notation 
and conventions of \cite{dWNsu8} and to second order in fermions is 
\begin{align}
 L= & -\frac{1}{2} R - \frac{1}{2} \overline{\Psi}_{M} \tilde{\Gamma}^{MNP} D_{N} \Psi_{P} 
 - \frac{1}{48} F^{MNPQ} F_{MNPQ} \notag \\[1mm]
   & - \,\frac{1}{(12)^3\sqrt{2}} i \epsilon^{MNPQRSTUVWX} F_{MNPQ} F_{RSTU} A_{VWX} 
   \notag \\[1mm]
  & - \, \frac{\sqrt{2}}{192} F_{MNPQ} \left(\overline{\Psi}_{R} \tilde{\Gamma}^{MNPQRS} \Psi_{S} + 12 \overline{\Psi}^{M} \tilde{\Gamma}^{NP} \Psi^{Q} \right),
\label{11dlag}
\end{align}
where the four-form field strength is
$$F_{MNPQ} = 4! \,\partial_{[M} A_{NPQ]}$$
and $D_{M}$ is the covariant derivative defined with respect to the metric 
\begin{equation}
 g_{MN} = E_{M}{}^{A} E_{N}{}^{B} \delta_{AB}. 
\end{equation}
The eleven-dimensional $\tilde{\Gamma}$ matrices~\footnote{We put a tilde in order to distinguish these $\Gamma$-matrices with
lower dimensional $\Gamma$-matrices to be introduced below, cf. (\ref{Gamma1}).} satisfy
\begin{equation}
 \{\tilde{\Gamma}_{A}, \tilde{\Gamma}_{B}\} = 2 \delta_{AB}, \qquad  \tilde{\Gamma}^{A_{1}\dots A_{11}} = -i \epsilon^{A_1 \dots A_{11} },
\end{equation}
where on the right hand side of the second equation we have suppressed a 
32-dimensional identity matrix.  
In this convention, the supersymmetry transformations of $D=11$ supergravity take the form 
\begin{align}
 \delta E_{M}{}^{A} &= \frac{1}{2} \overline{\varepsilon} \tilde{\Gamma}^{A} \Psi_{M}, 
 \label{susye} \\[1mm]
 \delta A_{MNP} &= - \frac{\sqrt{2}}{8} \overline{\varepsilon} \tilde{\Gamma}_{[MN} \Psi_{P]}, \label{susyA3}\\[1mm]
 \delta \Psi_{M} &= D_{M} \varepsilon + \frac{\sqrt{2}}{288} \left( \tilde{\Gamma}_{M}{}^{ABCD} - 8 E_{M}{}^{A} \tilde{\Gamma}^{BCD} \right) \varepsilon F_{ABCD}. \label{susypsi}
\end{align}

The appearance of exceptional global symmetries \cite{cremmerjulia} upon the toroidal reduction of the eleven-dimensional theory to dimensions $D \le 6$ requires the Hodge dualisation of all field strengths whose degrees are greater than or equal to $\textstyle{\frac{1}{2}}D.$  It should be emphasised \cite{CJLP} that this is a particular choice that is designed to maximise the global symmetry obtained under reduction.  Other dualisations will lead to other global symmetries in the reduced theory.  One can understand this choice by noting that the most obvious way in which the enhanced symmetry in the reduced theory manifests itself is the observation that the scalars in the reduced theory parametrise a coset whose numerator is the global symmetry group, while the denominator is the local symmetry group, which, in general, corresponds to the maximal compact subgroup of the global group.  In the reduction to dimensions $D \ge 6$, the scalar sector is clear and cannot be changed by a process of dualisation.  
However, this is 
not true for $D \le 5$, where one can increase the number of scalars by dualising higher degree field strengths.  Maximising the number of scalars in the reduced theory maximises the global symmetry obtained under reduction \cite{CJLP}.

The requirement for the dualisation of certain fields in the reduced theory for the manifestation of a larger symmetry group can be understood from an eleven-dimensional perspective by the need to include dual fields in eleven dimensions.  Thus, there seems to be an intimate connection between dualisation of fields in the reduced theory and dualisation in the full eleven-dimensional theory.  Indeed, this relation is explicitly demonstrated in Ref.\ \cite{GGPE8}, where the bosonic sector of the eleven-dimensional theory is reduced to three dimensions.  In the process of writing the scalar sector of the reduced theory as a non-linear coset sigma model with coset E$_{8(8)}$/SO$(16)$, one finds a precise relation between the three-dimensional dual scalar fields $\phi_m$ and $\psi^{mn}$ associated with the graviphoton ${B_{\mu}}^{m}$ and the three-form component $A_{\mu m n}$, respectively, and the purported eleven-dimensional duals $h_{m_1\ldots m_8,n}$ and $A_{m_1 \ldots m_6}$, respectively.  Therefore, given 
that 
our aim here is to understand the role of the four-dimensional global symmetry group E$_{7(7)}$ in eleven dimensions, it is natural that we should consider the dualisation of eleven-dimensional fields.

The dualisation of a form field can be understood on-shell as simply the Hodge dualisation of the field strength of the form field. However, the dualisation of gravity poses a difficult challenge. Meanwhile, the need for the dualisation of gravity is apparent not only from the perspective of the discussion in this paper, and other papers concerning the higher-dimensional origins of duality symmetries, but also from the fact that $D=11$ supergravity has solutions, such as the Kaluza-Klein monopole, that are expected to source the dual gravity field \cite{hulldual}.
Nevertheless, the elevation of gravitational duality to the non-linear level 
encounters a no-go theorem \cite{BBH}, which can only be evaded by a loss of either locality or Lorentz invariance. However, what is pertinent in this paper is the coupling of gravity to matter, in particular the 3-form of $D=11$ supergravity and its 6-form dual, and, moreover, the supersymmetry transformation of a candidate dual gravity field. In this case, the dualisation of gravity becomes problematic even at the linearised level \cite{BdRKKR}. The supersymmetry transformations can be made to close in the presence of a linearised dual gravity field if one takes a linearised approximation where one has only {\em global} supersymmetry. However, the supersymmetry transformation is no longer consistent with the equations of motion, or 
in other words the dual graviton does not carry the same degrees of freedom as the graviton even in a flat background \cite{BdRKKR}. Furthermore, it is argued in \cite{BdRKKR} that, under assumptions of locality and Lorentz invariance, it is not possible to dualise a linearised graviton field coupled to matter. Nevertheless, we find that the completion of the E$_{7(7)}$ matrix in eleven dimensions requires the existence of a field with the same representation as a dual gravity field. Moreover, we explicitly give the supersymmetry transformation of this field up to an undetermined constant in section \ref{sec:gengeoe7}. We should stress that our results are not in conflict with the no-go theorems of \cite{BBH, BdRKKR} as we will become apparent later. 

\subsection{Dualisation of the three-form potential} \label{sec:6formsusy}

The relevance of a six-form potential dual to the three-form potential within the context of $D=11$ supergravity was discussed soon after the eleven-dimensional theory was found \cite{NicTownsvN, DAuriaFre}.  Later, however, it was argued \cite{TownsDemoc} that such a potential is to be thought of as being sourced by a non-perturbative object---M5-brane---in a conjectured M-theory that goes beyond the supergravity theory.  As such, the incorporation \cite{deAlwis, BBS} of a six-form potential in the eleven-dimensional theory, including its supersymmetry transformation \cite{BBS} (see also \cite{NicTownsvN}), has been considered previously.  

Our interest in the six-form potential in this work will be limited to the form of its supersymmetry transformation.  The six-form potential dual of the three-form potential is introduced by considering its equation of motion, which can be simply derived from lagrangian \eqref{11dlag}:
\begin{equation}
 d F_{(7)} = \frac{7!\sqrt{2}}{2} \; F_{(4)} \wedge F_{(4)} -\frac{\sqrt{2}}{8} d \star X,
\end{equation}
where
\begin{equation}
 F_{(7)} = \star F_{(4)}
\end{equation}
and
\begin{equation}
 X^{MNPQ}=\overline{\Psi}_{R} \tilde{\Gamma}^{MNPQRS} \Psi_{S} + 12 \overline{\Psi}^{M} \tilde{\Gamma}^{NP} \Psi^{Q}.
\end{equation}
Observing that
\begin{equation}
 F_{(4)} \wedge F_{(4)} = \frac{3!}{7!} \; d(A_{(3)} \wedge F_{(4)})
\end{equation}
gives
\begin{equation}
 d \left( F_{(7)} - 3 \sqrt{2} \; A_{(3)} \wedge F_{(4)} + \frac{\sqrt{2}}{8} \star X \right)=0.
\end{equation}
Hence, there exists locally a six-form potential $A_{(6)}$ such that
\begin{equation}
 F_{(7)} = d A_{(6)} + 3 \sqrt{2} \; A_{(3)} \wedge F_{(4)} - \frac{\sqrt{2}}{8} \star X.
\end{equation}
Equivalently, in terms of components
\begin{align} \label{F7}
 F_{M_1 \dots M_7} & = 7! D_{[M_1} A_{M_2 \dots M_7]} + 7! \frac{\sqrt{2}}{2}  A_{[M_1 \dots M_3} D_{M_4} A_{M_5 \dots M_7]} \notag \\
 &\hspace*{40mm}- \frac{\sqrt{2}}{192} i \epsilon_{M_1 \dots M_{11}} \left(\overline{\Psi}_{R} \tilde{\Gamma}^{M_8 \dots M_{11} RS} \Psi_{S} + 12 \overline{\Psi}^{M_8} \tilde{\Gamma}^{M_9 M_{10}} \Psi^{M_{11}} \right).
\end{align}

As should be familiar to the reader, in this process we have interchanged the 
equations of motion and Bianchi identities.  Thus, the Bianchi identity satisfied by 
$F_{(7)}$ is equivalent to the equation of motion of $A_{(3)}$.  For our applications 
it is best to think of the above equation as a definition of potential $A_{(6)}$ in terms of 
the usual eleven-dimensional fields.  Therefrom, we can find the supersymmetry 
transformation of $A_{(6)}$.

Let us begin with an ansatz of the form
\begin{equation} \label{susygenA6}
  \delta A_{M_1\dots M_6} = \alpha \; \overline{\varepsilon} \tilde{\Gamma}_{[M_1 \dots M_5} \Psi_{M_6]} + \beta \; \overline{\varepsilon} \tilde{\Gamma}_{[M_1 M_2} \Psi_{M_3} A_{M_4 M_5 M_6]}.
\end{equation}
Now consider a supersymmetry variation of equation \eqref{F7}.  To fix the coefficients 
it suffices to consider terms of the form $D_{M} \epsilon$.  Hence, we concentrate on such terms:
\begin{align}
  i & {\epsilon_{M_1 \ldots M_7}}^{N_1 \ldots N_4} D_{N_1} \delta A_{N_2 N_3 N_4} - \frac{7!\sqrt{2}}{2} \; A_{[M_1 M_2 M_3} D_{M_4} \delta A_{M_5 M_6 M_7]} -7! \, D_{[M_1} \delta A_{M_2 \ldots M_7]}
\notag \\[3mm]
& \hspace{30mm} + \frac{\sqrt{2}}{96} i \epsilon_{M_1 \dots M_{11}} \left(\delta \overline{\Psi}_{R} \tilde{\Gamma}^{M_8 \dots M_{11} RS} \Psi_{S} + 12 \delta \overline{\Psi}^{M_8} \tilde{\Gamma}^{M_9 M_{10}} \Psi^{M_{11}} \right) + \dots = 0, \label{susytranF7}
\end{align}
where we have used the relation
\begin{equation}
 \overline{\psi} \, \tilde{\Gamma}^{A_1} \cdots \tilde{\Gamma}^{A_n} \chi = (-1)^n \ \overline{\chi} \, \tilde{\Gamma}^{A_n} \cdots \tilde{\Gamma}^{A_1} \psi.
\end{equation}
Substituting the supersymmetry transformations of the relevant fields using equations \eqref{susyA3}, \eqref{susypsi} and \eqref{susygenA6} into equation \eqref{susytranF7} gives
\begin{align}
7!(1/8-\beta) D_{[M_1} \overline{\varepsilon} & \tilde{\Gamma}_{M_2 M_3} \Psi_{M_4}  
A_{M_5 M_6 M_7]} \notag \\
& - 7! \alpha D_{[M_1} \overline{\varepsilon} \tilde{\Gamma}_{M_2 \ldots M_6} 
\Psi_{M_7]}  + \frac{\sqrt{2}}{96} i \, \epsilon_{M_1 \ldots M_{11}} D_{P} \overline{\varepsilon} \tilde{\Gamma}^{M_8 \ldots M_{11} P Q} \Psi_{Q} + \dots = 0.
\end{align}
Using the fact that
\begin{equation}
 \tilde{\Gamma}^{P_1 \ldots P_6} = -i/5! \ \epsilon^{P_1 \ldots P_6 Q_1 \ldots Q_5} \tilde{\Gamma}_{Q_1 \ldots Q_5}
\end{equation}
the above equation simplifies to
\begin{equation}
\left(\beta -\frac{1}{8} \right) D_{[M_1} \overline{\varepsilon} \tilde{\Gamma}_{M_2 M_3} \Psi_{M_4}  A_{M_5 M_6 M_7]}
+ \left(\alpha + \frac{3}{6!\sqrt{2}} \right) D_{[M_1} \overline{\varepsilon} \tilde{\Gamma}_{M_2 \ldots M_6} \Psi_{M_7]} + \dots = 0.
\end{equation}
Hence, the supersymmetry transformation of the 6-form dual is
\begin{equation}
  \delta A_{M_1\dots M_6} = - \frac{3}{6!\sqrt{2}} \overline{\varepsilon} \tilde{\Gamma}_{[M_1 \dots M_5} \Psi_{M_6]} + \frac{1}{8}  \overline{\varepsilon} \tilde{\Gamma}_{[M_1 M_2} \Psi_{M_3} A_{M_4 M_5 M_6]}. \label{susyA6}
\end{equation}
A complete proof of the consistency of this relation with transformations 
\eqref{susye}--\eqref{susypsi} requires
use of the Rarita-Schwinger equation for $\Psi_M$. For this reason, and because
duality can anyway be implemented only at the level of the equations of motion, the above 
supersymmetry transformation rules are jointly valid {\em on-shell} only. Nevertheless we should 
emphasise that, apart from this restriction, all formulae are valid at the full non-linear 
level, that is, we can simultaneously incorporate the 3-form and the 6-form into the
full $D=11$ theory.

\subsection{Dualisation of gravity} \label{sec:dualgrav}

Unlike the dualisation of the $3$-form potential of $D=11$ supergravity, 
the dualisation of gravity is only possible at the linearised level, where 
the eleven-dimensional metric is expanded according to
\begin{equation}\label{DGrav}
g_{MN} = \eta_{MN} +  h_{MN} + {\cal O}(h^2).
\end{equation}
In linearised general relativity, the dual graviton can either be formulated from 
Hodge dualising the Riemann two-form \cite{curtright, hulldual} or by Hodge dualising 
an index of the Einstein tensor \cite{weste11}. The generalisation of either approach 
at the non-linear level is obstructed by a no-go theorem \cite{BBH}, which can only 
be evaded (if it can be evaded at all) by abandoning either locality or Lorentz invariance or both. 
As shown in previous work \cite{curtright, dualHull1, weste11, hulldual} (see also \cite{Obers:1997kk,Obers:1998fb}), the field formally dual to the linearised metric  $h_{MN}$ is a 
mixed symmetry tensor $h_{M_1\dots M_8|N}$ that belongs to the (8,1) representation of  GL$(11,{\mathbb{R}})$ 
(with Dynkin label [1000000100]) and obeys the constraint
\begin{equation}\label{YTC}
h_{[M_1\dots M_8|N]} = 0.
\end{equation}
The dual graviton field thus belongs to a non-trivial Young tableau representation, 
and this feature is one main source of difficulty. At the linear level the gravitational analog 
of equation (\ref{F7}) is
\begin{equation}
Y_{M_1 \dots M_9|N} = 9! \, \partial_{[M_1} h_{M_2\dots M_9]|N},
\end{equation}
where
\begin{equation}\label{Y}
Y_{M_1\dots M_9|N} = \frac12 \epsilon_{M_1\dots M_9}{}^{PQ} \omega_{N\,PQ}
\end{equation}
with the associated linearised spin connection $\omega_{M\, NP} = 
2 \kappa \partial_{[N} h_{P]M}.$  Note that we do not need to distinguish 
between curved and flat indices since we are working to linear order.  The irreducibility 
constraint (\ref{YTC}) is equivalent to $\omega_M{}^M{}_N = 0$.~\footnote{The constraint 
  \eqref{YTC} appears naturally in all approaches based on E$_{10}$ \cite{DHN} and 
  E$_{11}$ \cite{weste11}, but one can also perform the dualisation without imposing it. 
  In this case \eqref{Y} must be replaced by 
  $$
  Y_{M_1\dots M_9|N} = \frac12 \epsilon_{M_1\dots M_9}{}^{PQ} \big(\omega_{N\,PQ}
             - 2\eta_{NP} \omega_R{}^R{}_Q\big)    \;\; .
 $$} 
It is now straightforward to show that the fields $h_{MN}$ and $h_{M_1\dots M_8|N}$
form a dual pair in the sense that the Bianchi identity for one implies the 
(linearised) equation of motion for the other.

As shown in \cite{BBH} it is not possible to elevate the duality relation (\ref{Y}) to the
interacting theory if one insists on locality and Lorentz invariance of the dual
formulation. These difficulties are also reflected in the impossibility of extending 
the duality between $h_{MN}$ and $h_{M_1\dots M_8|N}$ to the 
incorporation of matter, even if gravity is kept linear \cite{BdRKKR}.
The question of extending the gravitational duality to supergravity was
studied in \cite{BdRKKR}, though in terms of a simpler example, as well as
in unpublished work of the same authors.  The most general ansatz for the supersymmetry 
variation of the dual graviton compatible with the constraint (\ref{YTC}) reads
\begin{equation}\label{susyH}
\delta h_{M_1\dots M_8|N} \, \propto \,
\overline{\varepsilon} \tilde{\Gamma}_{M_1\dots M_8} \Psi_N -
\, \overline{\varepsilon} \tilde{\Gamma}_{N[M_1\dots M_7} \Psi_{M_8]} - \, 
C_0 \eta_{N[M_1} \overline{\varepsilon}  \tilde{\Gamma}_{M_2\dots M_7} \Psi_{M_8]} 
\end{equation}
with an undetermined constant $C_0$. For $C_0\neq 0$, the last term on the right hand side
leads to a breaking of GL$(11,{\mathbb{R}})$ covariance to SO(1,10).
It is clear that further restrictions must be imposed at this point. In particular,
we must restrict to {\em global supersymmetry} ($\partial_M\varepsilon = 0$) from 
the outset, otherwise the supersymmetry algebra cannot close on $h_{M_1\dots M_8|N}$ 
even at the linearised level.  This is easily seen by noting that the putative parameter 
\begin{equation}
\Lambda_{M_1\dots M_8} = \overline{\varepsilon}_1
\tilde{\Gamma}_{M_1\dots M_8}\varepsilon_2,
\end{equation}
which would be one of the gauge transformation parameters associated with the field
$h_{M_1\dots M_8|N}$ \cite{BBH} is {\em symmetric} under interchange of
$\varepsilon_1$ and $\varepsilon_2$, hence it cannot appear in the commutator of 
two local supersymmetries; instead, the commutator would lead to new 
transformations that cannot be interpreted as gauge transformations in the 
sense of \cite{BBH}. As a consequence it does not appear possible even at the
linearised level in this ``dual supergravity'' to consistently incorporate the 
gauge transformations necessary to remove unphysical helicity degrees of freedom.
One can nevertheless investigate the 
closure of the {\em global} supersymmetry algebra, which yields the
value $C_0 = 63/2$~\footnote{A.~Kleinschmidt, private communication.}. This calculation 
requires the consideration of  both the $\partial h \varepsilon$ and $F\varepsilon$ 
contributions in $\delta\Psi_N$. As we will see this is not the value we find here, cf.
(\ref{susyh}) below.

The difficulties outlined above, in our view, point to the core problem of properly
understanding the duality symmetries beyond their explicit realisation in dimensionally
reduced maximal supergravity: that is, the problem of {\em dualising Einstein's theory
at the non-linear level}. This is another indication that a proper understanding of
M-theory and the role of $D=11$ supergravity in this context will require the abandonment 
of conventional notions of covariance and space-time.

\section{Generalised geometry from eleven dimensions} \label{sec:e711d}

In this section we demonstrate by explicit construction how the bosonic degrees 
of freedom of $D=11$ supergravity can be assembled into E$_{7(7)}$-valued objects.
In particular the vector degrees of freedom can be combined into a $\bf{56}$-plet of E$_{7(7)}$, 
and the scalar fields into an E$_{7(7)}$-valued  56-bein $\cV$ in eleven dimensions, 
thus completing the construction of \cite{dWNsu8}. These results finally establish 
the relation between the old work of \cite{dWNsu8} with more recent constructions, 
where the existence of a generalised vielbein is usually postulated {\em ad hoc}
(usually with further constraints). They also link up with the original construction performed
in \cite{cremmerjulia} for the $T^7$ truncation of $D=11$ supergravity, but with the 
crucial difference that the present results are valid {\em in eleven dimensions.}

\subsection{SU(8) reformulation of $D=11$ supergravity}

In Ref.\ \cite{dWNsu8}, the eleven-dimensional theory is formulated in a manifestly SU(8) invariant manner. 
The eleven-dimensional space-time is split into a four-dimensional space-time and a seven-dimensional space. Hence the eleven-dimensional space-time coordinates and tangent coordinates are split as 
 \begin{equation}
  z^{M}= (x^{\mu}, y^{m}), \qquad z^{A}= (x^{\alpha}, y^{a}),
 \end{equation}
respectively. Furthermore, in an upper triangular gauge the elfbein takes the form 
\begin{equation} \label{elfbein}
 {E_M}^A = \begin{pmatrix}
                 \Delta^{-1/2} {e'_{\mu}}{}^{\alpha} & {B_{\mu}}^m {e_m}^a \\ 0 & {e_m}^a
                \end{pmatrix},
\end{equation}
where
\begin{equation}
 \Delta=\textup{det}({e_m}^a).
\end{equation}
Correspondingly, the eleven-dimensional gamma-matrices are decomposed in the following way
\begin{equation}\label{Gamma1}
 \tilde{\Gamma}_{\alpha}=\gamma_{\alpha} \otimes {\bf 1}, \qquad \tilde{\Gamma}_{a}=\gamma_{5} \otimes \Gamma_{a},
\end{equation}
where $\gamma_\alpha$ and $\Gamma_a$ satisfy the four and seven-dimensional Clifford algebras, respectively, and $${\gamma_5=\gamma_1\gamma_2\gamma_3\gamma_4}.$$ In particular,
\begin{equation}
 \Gamma^{a_1 \ldots a_7} = -i \epsilon^{a_1 \ldots a_7} {\bf 1}.
\end{equation}

The essence of the SU(8) invariant reformulation of the theory is in defining new fields 
with chiral SU(8) indices \cite{cremmerjulia,dWNsu8}~\footnote{For these we will use 
capital Roman letters $A,B,...$,
 the same as for flat indices in eleven dimensions. There should nevertheless arise no
 confusion as it should be clear from the context which kind of index is meant.}
\begin{gather}
 \varphi_\mu{}^A=\textstyle{\frac{1}{2}}(1+\gamma_5)\, \varphi'_\mu{}_{\bar{A}}, \qquad 
 \varphi_{\mu\, A}=\textstyle{\frac{1}{2}}(1-\gamma_5)\, \varphi'_\mu{}_{\bar{A}}, \\
 \chi^{ABC}= (1+\gamma_5) \chi'_{\bar{A}\bar{B}\bar{C}}, \qquad
 \chi_{ABC}= (1-\gamma_5) \chi'_{\bar{A}\bar{B}\bar{C}},
\end{gather}
where the indices on the right hand side are denoted with a bar to emphasise the fact that 
they are not chiral SU(8) indices, but SO(7) spinor indices.  We shall not make this distinction where the index type is clear from the context.  Fields $\varphi'$ and $\chi'$ are related to the original fields in the following manner~\footnote{Note that $(i \gamma_5)^{1/2}= \textstyle{\frac{1}{\sqrt{2}}}(1+i\gamma_5)$, while $(i \gamma_5)^{-1/2}= \textstyle{\frac{1}{\sqrt{2}}}(1-i\gamma_5)$.}
\begin{gather}
 \varphi'_\mu{} = \Delta^{-1/4} (i \gamma_5)^{-1/2} {e'_{\mu}}{}^{\alpha} (\Psi_{\alpha} - \textstyle{\frac{1}{2}}\gamma_5 \gamma_\alpha \Gamma^a \Psi_a), \\[3mm]
 \chi'_{ABC}=\textstyle{\frac{3}{4}} \sqrt{2} i \Delta^{-1/4} (i \gamma_5)^{-1/2} \Psi_{a[A} \Gamma^a_{BC]} .
\end{gather}
Similarly, the supersymmetry transformation parameter is redefined as follows
\begin{equation}
  \epsilon^A=\textstyle{\frac{1}{2}}(1+\gamma_5)\, \Delta^{1/4} (i \gamma_5)^{-1/2} \epsilon_{\bar{A}}, \qquad 
  \epsilon_A=\textstyle{\frac{1}{2}}(1-\gamma_5)\, \Delta^{1/4} (i \gamma_5)^{-1/2} \epsilon_{\bar{A}}.
\end{equation}
For the spin two degrees of freedom it then follows directly that
\begin{equation}
\delta e'_\mu{}^\alpha = 
\frac12 \overline{\epsilon}^A \gamma^\alpha \varphi_{\mu A} \; + \; {\rm h.c.}.
\end{equation}

\subsection{New generalised vielbeine} \label{sec:newgenv}
 
In the formulation of \cite{dWNsu8}, the local SU(8) symmetry is an enlargement of the local SO(7) symmetry of the tangent space of the seven-dimensional space. As such all fields with SO(7) tangent space indices are replaced in the reformulated theory with new fields carrying SU(8) indices. An important example of this is the generalised vielbein 
 \begin{equation}
  e^{m}_{AB} = i\Delta^{-1/2} \Gamma^m_{AB} \equiv 
  i \Delta^{-1/2} {e^m}_{a} \Gamma^{a}_{AB},
\label{gv1}
 \end{equation}
which replaces the siebenbein in the reformulated theory. This generalised vielbein 
is found \cite{dWNsu8} by considering the supersymmetry transformation:
\begin{equation} \label{susyB}
 \delta B_{\mu}{}^{m} = \frac{\sqrt{2}}{8} e^{m}_{AB} \left[ 2 \sqrt{2} \overline{\varepsilon}^{A} \varphi_{\mu}^{B} + \overline{\varepsilon}_{C} \gamma'_{\mu} \chi^{ABC} \right] 
 \, + \, \textup{h.c.}
\end{equation}
with $\gamma'_\mu \equiv e'_\mu{}^\alpha \gamma_\alpha$. Recently, it was found \cite{dWN13} 
that the supersymmetry transformation of a component of the 3-form $A,$
$$B_{\mu m n } = A_{\mu m n } - B_{\mu}{}^{p} A_{pmn}$$
leads to another generalised vielbein:
\begin{equation} \label{susyB3}
 \delta B_{\mu m n} = \frac{\sqrt{2}}{8} e_{mn}{}_{AB} \left[ 2 \sqrt{2} \overline{\varepsilon}^{A} \varphi_{\mu}^{B} + \overline{\varepsilon}_{C} \gamma'_{\mu} \chi^{ABC} \right] + \textup{h.c.}, 
\end{equation}
 where 
 \begin{equation}
  e_{mn}{}_{AB} = - \frac{\sqrt{2}}{12} i \Delta^{-1/2} \left( \Gamma_{mn}{}_{AB} + 6 \sqrt{2} A_{mnp} \Gamma^{p}_{AB} \right) 
\label{gv2}
 \end{equation}
with $\Gamma_{mn}\equiv e_m{}^a e_n{}^b \Gamma_{ab}$. Importantly, both generalised 
vielbeine transform in the same way under a supersymmetry transformation,
\begin{align}
 \delta e^{m}_{AB} &= - \sqrt{2} \Sigma_{ABCD} e^{m}{}^{CD} - 2 \Lambda^{C}{}_{[A} e^{m}_{B]C},
  \\[2mm]
 \delta e_{mn}{}_{AB} &= - \sqrt{2} \Sigma_{ABCD} e_{mn}{}^{CD} - 2 \Lambda^{C}{}_{[A} e_{mn}{}_{B]C}
\end{align}
with the complex self-dual SU(8) tensor
\begin{equation}
 \Sigma_{ABCD} = \bar{\varepsilon}_{[A} \chi_{BCD]} + \textstyle{\frac{1}{4!}} \epsilon_{ABCDEFGH} \bar{\varepsilon}^{E} \chi^{FGH}, 
\end{equation}
and where
\begin{equation}
 {\Lambda^B}_{A}= \textstyle{\frac{1}{8}} \, \bar{\varepsilon} \gamma_5 \Gamma_{ab} \Psi^{a} \; \Gamma^{b}_{AB} + 
 \textstyle{\frac{1}{8}} \, \bar{\varepsilon} \gamma_5 \Gamma_{a} \Psi_{b} \; \Gamma^{ab}_{AB} + 
 \textstyle{\frac{1}{16}} \, \bar{\varepsilon} \Gamma_{ab} \Psi_{c} \; \Gamma^{abc}_{AB}
\end{equation}
parametrises a field dependent local SU(8) rotation in eleven dimensions.

The generalised vielbeine $e^m_{AB}$ and $e_{mnAB}$ give rise to non-linear ans\"atze for the internal metric 
\cite{dWNW} and flux \cite{dWN13}, which pass some very non-trivial tests \cite{GGN}. The ans\"atze are obtained by comparing the supersymmetry transformations that lead to these vielbeine, \eqref{susyB} and 
\eqref{susyB3}, with the four-dimensional gauged supergravity supersymmetry 
transformations \cite{dWNN8, dWSTmax4}~\footnote{See also equation (7.10) in \cite{cremmerjulia}.}
\begin{align}
 \delta {A_{\mu}}^{IJ} &= - \textstyle{\frac{1}{2}} ({{u_{ij}}^{IJ}}+v_{ijIJ}) \left[ 2 \sqrt{2} \overline{\varepsilon}^{i} \varphi_{\mu}^{j} + \overline{\varepsilon}_{k} \gamma'_{\mu} \chi^{ijk} \right] 
 \, + \, \textup{h.c.}, \label{susyev} \\
 \delta A_{\mu IJ} &= - \textstyle{\frac{1}{2}} i({{u_{ij}}^{IJ}}-v_{ijIJ}) \left[ 2 \sqrt{2} \overline{\varepsilon}^{i} \varphi_{\mu}^{j} + \overline{\varepsilon}_{k} \gamma'_{\mu} \chi^{ijk} \right] 
 \, + \, \textup{h.c.},  \label{susymv}
\end{align}
for the 28 electric vectors $A_\mu{}^{IJ}$ and the 28 magnetic vectors $A_{\mu IJ}$, respectively.
Here $i,j,k, \ldots$ are SU$(8)$ indices, while $I,J,K, \ldots$ are SL$(8,{\mathbb{R}})$ 
indices (which are to be considered as SO(8) indices after gauging).  Moreover, since 
the linear Kaluza-Klein ansatz for vector fields is exact, $B_{\mu}{}^{m}$ and $B_{\mu mn}$ are related to ${A_{\mu}}^{IJ}$ and ${A_{\mu IJ}},$ respectively via the 28 $S^7$ Killing vectors $K^{mIJ}(y)$ and 28 two-forms 
$
K^{mnIJ}= \eo_a{}^m \eo_b{}^n\,
\overline{\eta}^{I} \Gamma^{ab} \eta^{J},
$
where $\eta^I$ are the $S^7$ Killing spinors, and $\eo_a{}^m$ is the inverse
siebenbein on $S^7$.  This gives a relation between the generalised vielbeine and the scalars of the four-dimensional gauged supergravity. The non-linear 
ans\"atze obtained in this way are highly non-trivial and there is no purely bosonic argument available to derive them otherwise. Indeed the non-linear metric ansatz is part of the 
consistency proof of the $S^7$ reduction \cite{dWNconsis}. Meanwhile, the recently discovered non-linear flux ansatz is shown \cite{GGN} to be an efficient way to analytically find the internal flux associated to not only a four-dimensional maximally gauged supergravity critical point, but even a whole family. 

The generalised vielbeine, \eqref{gv1} and \eqref{gv2}, were found by considering the supersymmetry transformation of fields that under reduction would correspond to vector fields, 
{\it viz.} \ $B_{\mu}{}^{m}$ and $B_{\mu mn}.$ In the maximally gauged theory these vector fields each give rise to 28 vector fields, accounting for the 56 vector fields of which 28 appear in the gauged theory lagrangian. However, in the ungauged theory these vector fields only account for 28 of the 56 vector fields. The other 28 vector fields are the duals of these fields in 4-dimensions. We can view these dual fields as coming from the reduction of fields that are the dualisations of the eleven-dimensional fields. Therefore, we next turn to the supersymmetry transformation 
of the 6-form potential in eleven dimensions that is dual to the 3-form gauge potential,
with the aim of extracting from it another set of vector components with an associated generalised vielbein. Consider the following components of the 6-form:
$$B_{\mu m_1 \dots m_5} = A_{\mu m_1 \dots m_5} - B_{\mu}{}^{p} A_{p m_1 \dots m_5}.$$ 
The supersymmetry transformation of the 6-form, equation \eqref{susyA6}, can now be used to show that
\begin{equation} \label{susyB6}
\delta \left( B_{\mu m_1 \dots m_5} - \frac{\sqrt{2}}{4} B_{\mu [m_1 m_2} A_{m_3 m_4 m_5]} \right) = \frac{\sqrt{2}}{8} e_{m_1 \dots m_5}{}_{AB} \left[ 2 \sqrt{2} \overline{\varepsilon}^{A} \varphi_{\mu}^{B} + \overline{\varepsilon}_{C} \gamma'_{\mu} \chi^{ABC} \right] + \textup{h.c.}
\end{equation}
with the associated new generalised vielbein  
\begin{align}
 e_{m_1 \dots m_5}{}_{AB}&= \frac{1}{6!\sqrt{2}} i \Delta^{-1/2} \Bigg[ \Gamma_{m_1 \dots m_5}{}_{AB} + 60 \sqrt{2} A_{[m_1 m_2 m_3} \Gamma_{m_4 m_5]}{}_{AB} \notag \\ 
 & \hspace*{50mm}  - 6! \sqrt{2} \Big( A_{p m_1 \dots m_5} -  \frac{\sqrt{2}}{4} A_{p[m_1 m_2} A_{m_3 m_4 m_5]} \Big) \Gamma^{p}_{AB} \Bigg]. \label{gv3}
\end{align}
This new vielbein depends not only on the metric and 3-form along the seven internal 
directions, but also on the 6-form potential $A_{m_1 \dots m_6}.$ 

Using the identities listed in appendix \ref{app:susytrans}, one can show that the supersymmetry transformation of this  generalised vielbein takes the same form as for the other generalised vielbeine, i.e.
\begin{equation}
  \delta e_{m_1 \dots m_5}{}_{AB}= - \sqrt{2} \Sigma_{ABCD} e_{m_1 \dots m_5}{}^{CD} - 2 \Lambda^{C}{}_{[A} e_{m_1 \dots m_5}{}_{B]C}.
\end{equation}

Remarkably all generalised vielbein components transform in exactly the same way
under local supersymmetry, and with the {\em same} compensating SU(8) rotation.
We emphasise again that all formulae are valid in eleven dimensions, and  at the 
full non-linear level. Furthermore at no point was it necessary to truncate or impose any 
restriction on the coordinate dependence. It is now straightforward to derive the 
non-linear ansatz for the 6-form field $A_{m_1\dots m_6}$ by substituting
the relevant expressions in terms of $S^7$ Killing vectors and the four-dimensional
fields on the left hand side of\eqref{gv3}, and then projecting out the last component on the right hand side.
A detailed discussion will, however, be given elsewhere.

\subsection{Generalised vielbeine and E$_{7(7)}$} \label{sec:gengeoe7}

The similarity of the transformations for the generalised vielbeine $e^m_{AB}\, ,\, e_{mnAB}$ 
and $e_{m_1\dots m_5AB}$ suggests that these are components of a {\em single} object
in {\em eleven} dimensions, namely a 56-bein
\begin{equation}\label{E7GV}
 \cV(z) \, \equiv \big(\cV^{\tM \tN}{}_{AB}(z),  \cV_{\tM \tN\, AB}(z)\big)
\; \in \; {\rm E}_{7(7)}/{\rm SU}(8)  
\end{equation}
and its complex conjugate $\big(\cV^{\tM \tN}{}_{AB} (z), \cV_{\tM \tN\, AB}(z)\big)^* \equiv
\big( \cV^{\tM \tN \,AB}(z), \cV_{\tM \tN}{}^{AB}(z)\big)$.  Indices $\tM , \tN = 1,...,8$
are associated with the SL(8,$\mathbb{R}$) subgroup of E$_{7(7)}$. Accordingly, we proceed
from the following identification of this new object with the generalised vielbeine
obtained so far~\footnote{The extra factor of $\Delta$ in the second line, 
  and in \eqref{cV2} below, is necessary in order to maintain the form 
  of the supersymmetry variation given in (\ref{susyV}).}
\begin{eqnarray}
\cV^{m8}{}_{AB} &\equiv& e^m_{AB}, \quad \; \quad \cV_{mn\, AB} \equiv e_{mnAB} \nonumber\\
\cV^{mn}{}_{AB} &\equiv& \frac1{5!} \Delta \epsilon^{mnp_1\dots p_5} e_{p_1\dots p_5 AB} \label{cV1}
\end{eqnarray}
in accordance with the decomposition
\begin{equation}
{\bf 56} \; \rightarrow \; {\bf 28} \oplus \overline{\bf 28} \; \rightarrow \; {\bf 7} \oplus {\bf 21} \oplus \overline{\bf 21} \oplus \overline{\bf 7}
\label{gl7decom}
\end{equation}
of the $\bf 56$ representation of E$_{7(7)}$ under its SL(8,$\mathbb{R}$) and 
GL(7,$\mathbb{R}$) subgroups. 
Dropping the compensating SU(8) rotation the supersymmetry variations 
obtained in the foregoing section are then all consistent with the formula
\begin{equation} \label{susyV}
\delta \cV^{\tM\tN}{}_{AB}  = - \sqrt{2} \Sigma_{ABCD} \cV^{\tM\tN\,CD}\; , \quad
\delta \cV_{\tM\tN\, AB}  = - \sqrt{2} \Sigma_{ABCD} \cV_{\tM\tN}{}^{CD}
\end{equation}
which upon reduction to four dimensions precisely coincides with the variation 
of the 56-bein in $N=8$ supergravity. Because the theory by construction is invariant
under local SU(8) in eleven dimensions, this confirms that the vielbein components 
identified up to here are indeed part of an E$_{7(7)}/$SU(8) coset element 
${\cal{V}}(z^M)$ {\em in eleven dimensions}. 

The 56 vectors can likewise be assembled into a single object of the form 
$({\mathcal{B}_{\mu}}^{\tM \tN}, \mathcal{B}_{\mu \, \tM \tN})$. With the identifications
obtained so far, we define
\begin{equation}\label{calB}
{\mathcal{B}_{\mu}}^{m8} \equiv {B_{\mu}}^{m}\;, \quad 
\mathcal{B}_{\mu \, mn} \equiv B_{\mu m n}\;  , \quad
{\mathcal{B}_{\mu}}^{mn} \equiv \frac{1}{5!} \Delta \epsilon^{m n p_1\dots p_5} 
\left( B_{\mu p_1 \dots p_5} - \frac{\sqrt{2}}{4} B_{\mu [p_1 p_2} A_{p_3 p_4 p_5]} \right).
\end{equation}
The remaining `missing' component
\begin{equation}\label{B7}
\mathcal{B}_{\mu \, m8} \equiv \frac{1}{7!} \Delta \epsilon^{n_1\dots n_7} 
\mathcal{B}_{\mu n_1 \ldots n_7,m}
\end{equation}
will be given in equation 
\eqref{Btilde} below.  Now, the results for the supersymmetry variations of the vectors 
introduced above can be summarised by the following simple transformation formulae
\begin{align} \label{susycalB}
 \delta {\mathcal{B}_{\mu}}^{\tM \tN} &= \frac{\sqrt{2}}{8} \cV^{\tM\tN}{}_{AB} \left[ 2 \sqrt{2} \overline{\varepsilon}^{A} \varphi_{\mu}^{B} + \overline{\varepsilon}_{C} \gamma'_{\mu} \chi^{ABC} \right] + \textup{h.c.}, \notag \\[3mm]
 \delta \mathcal{B}_{\mu \, \tM \tN} &= \frac{\sqrt{2}}{8} \cV_{\tM\tN \, AB} \left[ 2 \sqrt{2} \overline{\varepsilon}^{A} \varphi_{\mu}^{B} + \overline{\varepsilon}_{C} \gamma'_{\mu} \chi^{ABC} \right] + \textup{h.c.}
\end{align}
complementing the supersymmetry transformations \eqref{susyV} of $\cV$. 
These transformations now have exactly the same form as the ones for the corresponding 
variations of the $D=4$ fields, but they are here valid in eleven dimensions. Note also
that the distribution of the 28 {\em physical} spin-one degrees of freedom between these
56 vectors depends on the given compactification. By comparing these with the
variations \eqref{susyev} and \eqref{susymv} and substituting the identifications
\eqref{cV1} we can now in principle derive non-linear ans\"atze for all $D=11$ fields
{\em and their duals}\,!

The last missing seven components \eqref{B7} corresponding to the $\overline{\bf 7}$ in  
the decomposition \eqref{gl7decom}, whose existence we had already anticipated above,  
turn out to be related, not unexpectedly, to dual gravity. In order to identify them 
and to complete the E$_{7(7)}$ matrix, we note that these components of the 
matrix ${\cal{V}}(z)$ must be of the form  
\begin{equation}
\cV_{m8\,AB} = \frac1{7!} \Delta \epsilon^{n_1\dots n_7} e_{n_1\dots n_7, m AB}. \label{cV2}
\end{equation}
A calculation now shows that the correct expression is given by
\begin{align}
 e_{m_1 \dots m_7, n}{}_{AB} &=- \frac{2}{9!} i \Delta^{-1/2} \Bigg[ (\Gamma_{m_1 \dots m_7} \Gamma_{n}{})_{AB} + 126 \sqrt{2}\ A_{n [m_1 m_2} \Gamma_{m_3 \dots m_7]}{}_{AB} \notag \\ 
 & \hspace*{25mm}  + 3\sqrt{2} \times 7! \Big( A_{n [ m_1 \dots m_5} + \frac{\sqrt{2}}{4} A_{n[m_1 m_2} A_{m_3 m_4 m_5} \Big) \Gamma_{m_6 m_7]}{}_{AB} \notag \\
  & \hspace*{25mm} + \frac{9!}{2} \Big(A_{n [ m_1 \dots m_5} + \frac{\sqrt{2}}{12} A_{n[m_1 m_2} A_{m_3 m_4 m_5} \Big) A_{m_6 m_7] p } \Gamma^{p}{}_{AB} \Bigg]. \label{gv4}
\end{align} 
This component is found by insisting that $e_{m_1 \dots m_7, n}{}_{AB}$ transform as 
\begin{equation} \label{susygenv4}
   \delta e_{m_1 \dots m_7, n}{}_{AB} = - \sqrt{2} \Sigma_{ABCD} e_{m_1 \dots m_7, n}{}^{CD} - 2 \Lambda^{C}{}_{[A} e_{m_1 \dots m_7, n}{}_{B]C},
\end{equation}
just as the other components of $\cV$. In this case the coefficients must take the specific 
values that appear in the definition \eqref{gv4}.~\footnote{Of course, an overall 
   rescaling by a real constant is allowed.} In other words, the form of the supersymmetry 
variation and the compensating SU(8) rotation uniquely fixes all coefficients.

In accordance with our previous findings we would expect this generalised vielbein
to come from the supersymmetry variation of the vector associated to the
 $(M_1\dots M_8|N)\equiv (\mu m_1\dots m_7| n)$ component of the 
dual gravity field $h_{M_1\dots M_8|N}$. Ignoring difficulties related to the non-linear
extension of duality in eleven dimensions, we find that indeed
the above generalised vielbein comes from the supersymmetry transformation of 
\begin{align}
 \mathcal{B}_{\mu m_1 \dots m_7, n} \equiv  B_{\mu m_1 \dots m_7, n} - B_{\mu [ m_1 \dots m_5} A_{m_6 m_7] n} + c \, 5! & (2 \sqrt{2}) B_{[\mu m_1 \dots m_5} B_{m_6 m_7] n} \notag \\
 & + \frac{\sqrt{2}}{12} B_{\mu [m_1 m_2} A_{m_3 \dots m_5} A_{m_6 m_7] n},
\label{Btilde}
\end{align}
if and only if the supersymmetry transformation of the new field $B_{\mu m_1\dots m_7,n}$
({\em not}  $\cB_{\mu m_1\dots m_7,n}$!) is
\begin{align}
 \delta B_{\mu m_1 \dots m_7, n} &= -\frac{1}{9!} \left( \overline{\varepsilon} \tilde{\Gamma}_{\mu m_1 \dots m_7} \Psi_{n} - 8 \overline{\varepsilon} \tilde{\Gamma}_{n}  \tilde{\Gamma}_{[\mu m_1 \dots m_6} \Psi_{m_7]} \right) + c \overline{\varepsilon} \tilde{\Gamma}_{[\mu m_1 \dots m_4} \Psi_{m_5} A_{m_6 m_7] n}  \notag \\
 & \quad + \frac{\sqrt{2}}{3} \overline{\varepsilon} \tilde{\Gamma}_{[\mu m_1} \Psi_{m_2} \left( A_{m_3 \dots m_7]n} + \frac{\sqrt{2}}{12} A_{m_3 \dots m_5} A_{m_6 m_7]n} \right)  \notag \\ 
 & \qquad -  c \, 5! \overline{\varepsilon} \tilde{\Gamma}_{[\mu m_1} \Psi_{m_2} \left( A_{m_3 \dots m_7]n} + \frac{\sqrt{2}}{4} A_{m_3 \dots m_5} A_{m_6 m_7]n} \right),  \label{susyh}
\end{align}
where $c$ is an undetermined constant. More specifically, we have
\begin{equation} \label{susyB81}
  \delta \mathcal{B}_{\mu m_1 \dots m_7, n} = 
  \frac{\sqrt{2}}{8}  e_{m_1 \dots m_7, n}{}_{AB} \left[ 2 \sqrt{2} \overline{\varepsilon}^{A} \varphi_{\mu}^{B} + \overline{\varepsilon}_{C} \gamma'_{\mu} \chi^{ABC} \right] + \textup{h.c.}.
\end{equation}
The indeterminacy encoded in the constant $c$ can be viewed as a consequence of
the fact that there is no contribution $\propto B_\mu{}^p h_{pm_1\dots m_7,n}$ in the
definition of  the field $\mathcal{B}_{\mu m_1 \dots m_7, n}$, unlike for the other 
components of the vector fields. In fact, the structure of the first two terms on the right hand side
of \eqref{susyh} is partly determined by requiring the absence of terms involving
$B_\mu{}^n$ in its variation under local supersymmetry.

We see that the first two terms on the right hand side of equation \eqref{susyh} 
disagree with the eleven-dimensional ansatz (\ref{susyH}), even though the representation 
constraint (\ref{YTC}) is trivially satisfied for all terms on the right hand side by virtue of Schouten's 
identity as applied to seven dimensions. Nevertheless, the above result is valid 
{\em at the full non-linear level}. Equally important, the supersymmetry algebra
is expected to close properly on-shell on all components of the 56-bein $\cV$, 
because our theory is physically equivalent  on-shell to the original $D=11$ supergravity
(although with a suitable re-interpretation of the symmetries).
There appears to be no immediate contradiction with the no-go
theorems of \cite{BBH, BdRKKR} because we have abandoned general covariance and Lorentz
invariance in eleven dimensions in the course of our construction. However, the disagreement does seem to suggest that the supersymmetry transformation \eqref{susyh} is only valid for the particular components given and is not to be regarded as part of a covariant expression for the supersymmetry transformation of a dual gravity field, at least not in a simple way.

Let us now return to the question of how these results relate to more recent studies
of generalised geometry. The central object there is an element of the 
duality coset under consideration and is usually also referred to as the ``generalised vielbein.''  This generalised vielbein, which {\it a priori} could be different from the one 
identified here is constructed using a non-linear realisation \cite{BO, west2000, locale11}, which is a group theoretic method for computing a coset element in a particular representation of the numerator group.  For the E$_{7(7)}$/SU$(8)$ duality coset, the non-linear realisation gives a coset element in the fundamental $\bf{56}$ representation of E$_{7(7)}$, that is uniquely decomposed under its GL(7,$\mathbb{R}$) subgroup as described in equation \eqref{gl7decom} \cite{hillmann}.  In order to compare this construction with the 56-bein derived here, rewrite the 56-bein components as follows
\begin{align}\label{triangV}
 \cV_{m8 \, AB} &= {\cV_{m8}}^{a} \Gamma_{a \, AB} + \cV_{m8 \, ab} \Gamma^{ab}_{AB} + i {\cV_{m8}}^{ab} \Gamma_{ab \, AB} + i \cV_{m8 \, a} \Gamma^{a}_{AB}, \notag\\
 {\cV^{mn}}_{AB} &= \hspace{21mm} {\cV^{mn}}_{ab} \Gamma^{ab}_{AB} + i \cV^{mn \, ab} \Gamma_{ab \, AB} + i {\cV^{mn}}_{a} \Gamma^{a}_{AB}, \notag\\
 \cV_{mn \, AB} &=\hspace{44mm} i {\cV_{mn}}^{ab} \Gamma_{ab \, AB} + i \cV_{mn \, a} \Gamma^{a}_{AB}, \notag\\
 {\cV^{m8}}_{AB} &= \hspace{71mm} i {\cV^{m8}}_{a} \Gamma^{a}_{AB},
\end{align}
where the precise coefficients of the $\Gamma$-matrices on the right hand side can be computed from the definition of $\cV$, equations \eqref{cV1} and \eqref{cV2}, and the definitions of the generalised vielbeine, equations \eqref{gv1}, \eqref{gv2}, \eqref{gv3} and \eqref{gv4}. 
Forming a $4\times 4$ block matrix with the coefficients on the right hand side as is 
suggested by the structure of the equations above one finds that the form of this matrix 
is precisely the same as that found in Ref.\ \cite{hillmann} (see the matrix labelled 
$\mathcal{R(V)}$ on the top of page 21 in Ref.\ \cite{hillmann}).
Of course, the precise numerical factors are different, but this is due to differing conventions.  What is important is the form of each element and the precise factors of $\Delta$, which agree. Furthermore, this matrix agrees with the E$_{7(7)}$/SU$(8)$ coset element also calculated by non-linear realisation in \cite{BGPW}, up to an overall $\Delta$ (equation (127) of Ref.\ \cite{BGPW}), which is due to the fact that in \cite{BGPW} the E$_{7(7)}$ algebra is taken to be embedded in E$_{11}$.   

While the triangular structure evident in \eqref{triangV} has been known for a long time
to emerge in the reduction to four dimensions \cite{cremmerjulia}, the new feature
here is that all relations displayed are now valid in eleven dimensions. In particular,
and as with the first two generalised vielbeine, by comparing transformations \eqref{susyB6} and \eqref{susyB81} to \eqref{susyev} and \eqref{susymv} one can now construct a 
non-linear ansatz also for the dual field $A_{m_1 \dots m_6}.$  As is demonstrated 
in appendix \ref{app:englert} for the Englert solution \cite{englert}, the six-form 
potential is expected to be generically non-zero for any compactification other than the 
torus reduction of \cite{cremmerjulia}. The new non-linear flux ansatz would, in principle, 
give $A_{m_1 \dots m_6}$ from the expectation values of the four-dimensional scalars.  
In particular, it would reproduce $A_{m_1 \dots m_6}$ of the Englert solution given 
in appendix \ref{app:englert}.   

\section{Generalised Vielbein Postulate} \label{sec:gvp}

In the SU$(8)$ invariant reformulation of $D=11$ supergravity the generalised 
vielbein $e^{m}_{AB}$ satisfies a number  of consistency relations, collectively
referred to as the \emph{generalised vielbein postulate}. These are  differential
relations for the action of the $D=11$ derivatives on the vielbeine. For the seven 
internal directions, they read
\begin{equation}\label{GVP0}
 \partial_{m} e^{n}_{AB} + \cQ_{m}^{C}{}_{[A} e^{n}_{B]C} + \cP_{mABCD} e^{nCD}= 0.
\end{equation}
The E$_{7(7)}$ connection coefficients $\cQ_{m}^{A}{}_{B}$ 
and $\cP_{mABCD}$ \footnote{These coefficients are denoted by $\cB_m^A{}_B$
  and $\cA_{m ABCD}$ in \cite{dWNsu8}.} are of the form
\begin{gather}
 \cQ_{m}^{A}{}_{B} = \textstyle{\frac{1}{2}} ({e^p}_{a} \partial_{m} e_{p\, b}) \Gamma^{ab}_{AB} + \textstyle{\frac{\sqrt{2}}{14}} i f e_{m a} \Gamma^{a}_{AB} - \textstyle{\frac{\sqrt{2}}{48}} e_{m}^{a} F_{abcd} \Gamma^{bcd}_{AB}, \\[3mm]
 \cP_{mABCD} = - \textstyle{\frac{3}{4}} ({e^p}_{a} \partial_{m} e_{p\, b}) \Gamma^{a}_{[AB} \Gamma^{b}_{CD]} +  \textstyle{\frac{\sqrt{2}}{56}} i f {e_{m}}^{a} {\Gamma_{ab}}_{[AB} \Gamma^{b}_{CD]} + \textstyle{\frac{\sqrt{2}}{32}} e_{m}^{a} F_{abcd} \Gamma^{b}_{[AB} \Gamma^{cd}_{CD]},
\end{gather}
where
\begin{equation}
f= -\textstyle{\frac{1}{24}} i \eta^{\alpha \beta \gamma \delta} F_{\alpha \beta \gamma \delta}.
\end{equation}
Note that the partial derivative $\partial_m$ can be traded for a background covariant 
derivative $\Do_m$ appropriate for the $S^7$ compactification as explained in \cite{dWNsu8}. Let us now take a look at how \eqref{GVP0} generalises to the new vielbein components identified in this paper. In doing so, we will not aim
for completeness, as much further work is obviously required to penetrate the 
structures exhibited here.

It takes a bit of algebra to check that, in fact, the new generalised vielbeine do satisfy analogous relations. More precisely, we find
\begin{gather}
   \partial_{p} e_{mn AB} + \Xi_{p|mn|q} e^{q}_{AB} + \cQ_{p}^{C}{}_{[A} e_{mn B]C} + 
   \cP_{pABCD} e_{mn}{}^{CD}= 0, \label{7gvp2} \\[5mm]
    \partial_{p} e_{m_1 \dots m_5 AB} + \Xi_{p|m_1 \dots m_5|q} e^{q}_{AB} + 
    \Xi_{p|m_1 \dots m_5}{}^{qr} e_{qrAB} + \cQ_{p}^{C}{}_{[A} e_{m_1 \dots m_5 B]C} 
    + \cP_{pABCD} e_{m_1 \dots m_5}{}^{CD}= 0, \\
    \hspace{-50mm} \partial_{p} e_{m_1 \dots m_7, n AB} + \Xi_{p|m_1 \dots m_7, n}{}^{qr} e_{qr AB} + 
    \Xi_{p|m_1 \dots m_7,n}{}^{q_1 \dots q_5} e_{q_1 \dots q_5 AB} \notag \\[2mm] \hspace{70mm} + \cQ_{p}^{C}{}_{[A} e_{m_1 \dots m_7, n B]C} 
    + \cP_{pABCD} e_{m_1 \dots m_7, n}{}^{CD}= 0, \label{7gvp4}
\end{gather}
where
\begin{align} \label{Xi1}
 \Xi_{p|mn|q} &\equiv\,  \partial_{p} A_{mnq} - \frac{1}{4!} F_{pmnq}, \\[1mm]
 \Xi_{p|m_1 \dots m_5|q} &\equiv \, \partial_{p} A_{q m_1 \dots m_5} + \frac{\sqrt{2}}{48} F_{p[qm_1 m_2} A_{m_3 \dots m_5]} \notag \\[3pt]
 & \qquad - \frac{\sqrt{2}}{2} \left( \partial_{p} A_{[q m_1 m_2} - \frac{1}{4!} F_{p[qm_1 m_2} \right) A_{m_3 \dots m_5]} - \frac{1}{7!} F_{pq m_1 \dots m_5}, \\[1mm]
  \Xi_{p|m_1 \dots m_5}{}^{qr} &\equiv \,  \frac{1}{\sqrt{2}} \Xi_{p|[m_1 m_2|m_3} \delta^{\,q\,r}_{m_4 m_5]}, \\[3mm]
  \Xi_{p|m_1 \dots m_7, n}{}^{qr} &\equiv \, - \Xi_{p|[m_1 \dots m_5||n|} \delta^{\,q\,r}_{m_6 m_7]}, \\[3mm]
  \Xi_{p|m_1 \dots m_7, n}{}^{q_1 \dots q_5} &\equiv \,  \Xi_{p|[m_1 m_2||n|} \delta^{q_1 \dots q_5}_{m_3 \dots m_7]}.
 \end{align}
As was to be expected from the explicit dependence of the new vielbein components on
the 3-form and 6-form potentials, there appear terms which are {\em not gauge invariant}.
However, closer inspection of these expressions now reveals a truly remarkable feature,
not at all obvious nor to be expected from \eqref{gv2}, \eqref{gv3} and \eqref{gv4}: they all vanish upon 
anti-symmetrisation, and therefore precisely correspond to the Young tableaux that 
are eliminated by projecting onto the gauge invariant field strengths upon acting with
a derivative on the 3-form or 6-form potential! More specifically,
we have
\begin{equation}
\partial_m A_{npq} \,=\, \frac1{4!} F_{mnpq} \, + \, \Xi_{m|np|q}
\end{equation}
corresponding to the Young tableau decomposition
\begin{equation}
\yng(1,1,1) \;\; \bigotimes \;\; \yng(1) \; = \; \yng(1,1,1,1) \;\;  \bigoplus \;\; \yng(2,1,1)
\end{equation}
and similarly for the 7-form field strength
\begin{equation}
\partial_{m_1} A_{m_2\dots m_7} \, = \, \frac1{7!} F_{m_1\dots m_7} \, + \,
             \Xi_{m_1|m_2\dots m_6|m_7}
\end{equation}
corresponding to 
\begin{equation}
\yng(1,1,1,1,1,1) \;\; \bigotimes \;\; \yng(1) \; = \; \yng(1,1,1,1,1,1,1) \;\; \bigoplus \;\; 
\yng(2,1,1,1,1,1).
\end{equation}

In Ref.\ \cite{dWNcc}, a version of the vielbein postulate was given {\em with} Christoffel
symbols along the internal directions included, so the above findings motivate a similar
interpretation of the $\Xi$ symbols as {\em generalised connections} along the 
remaining directions of the E$_{7(7)}$ vielbein \eqref{E7GV}, in accordance with the decomposition \eqref{gl7decom}. More precisely, this symbol would be of the form
$\Xi^{\tM\tN}_{\tP\tQ\, \tR\tS}$, where the SL(8,$\mathbb{R}$) index pairs can appear either 
in the upper or the lower position. Because the gauge invariant field strengths are
part of the connection coefficients $\cQ_m^A{}_B$ and $\cP_{m ABCD}$ the above
decompositions should thus be regarded on a par with the corresponding decomposition of the usual vielbein derivative, {\it viz.}
\begin{equation}
\partial_M E_N{}^A \, = \, \omega_{MN}{}^A \,+\, \Gamma_{MN}{}^A
\end{equation}
into a piece covariant with respect to general coordinate transformations, and a non-gauge
invariant piece, thus extending ordinary geometry so as to comprise the
$p$-form fields of $D=11$ supergravity. This interpretation is further supported 
by the fact that, like the standard connections, the above objects are
{\em not} gauge invariant under the respective 2-form and 5-form gauge transformations,
in line with the interpretation of the latter as new coordinate transformations, while the
non-gauge covariant part of the variation drops out in the difference of two such
connections, again in complete analogy with usual affine connections. 

Moreover, note that the generalised vielbein $e^{m}_{AB}$ is absent from the generalised vielbein postulate for $e_{m_1 \dots m_7, n AB},$ equation \eqref{7gvp4}, or equivalently 
\begin{equation}
 \Xi_{p|m_1 \dots m_7, n |q} \equiv 0.
\end{equation}
For this term to be non-zero, it would be required for it to contain undifferentiated 3-form or 6-form potentials, which would introduce non-gauge invariances beyond what would be expected from connection components. Therefore, the vanishing of the above term is desirable from this perspective. 

The appearance of non-gauge invariant expressions for the 3-form and 6-form
gauge fields may appear strange at first sight, because all investigations of their
role in supergravity and superstring theory have so far focused exclusively on the
gauge invariant $(p+1)$-form field strengths. In this regard it is noteworthy that
the level expansion of the E$_{10}$ algebra gives rise to an infinite tower of
so-called `gradient representations', which have been tentatively associated with the
(time derivative of the) spatial gradients of the 3-from and 6-form fields \cite{DHN},
and where there is likewise no anti-symmetrisation in the spatial indices. In string
theory, the gauge invariant field strengths are associated to $D(p-1)$-branes, widely
considered a key ingredient towards a better understanding of non-perturbative
string theory. Our partial results above again underline the necessity of coming to
grips with {\em non-trivial Young tableau representations}, which in the level expansion
of E$_{10}$ constitute the vast majority of representations \cite{NFlevel}.

For the derivatives along the space-time directions, we have similar relations
which now also involve the vector fields $B_\mu{}^m \, , \, B_{\mu mn}$ and 
$B_{\mu m_1\dots m_5}$.  The components $e^{m}_{AB}$ are already known 
to satisfy the following equation \cite{dWNsu8}
\begin{equation}
 \cD_\mu e^{m}_{AB} + \frac{1}{2} \partial_{n} B_{\mu}{}^{n} e^{m}_{AB} + \partial_{n} B_{\mu}{}^{m} e^{n}_{AB} + \cQ_{\mu}^{C}{}_{[A} e^{m}_{B]C} + \cP_{\mu ABCD} e^{mCD} = 0, 
\end{equation}
where $\cD_\mu \equiv \partial_\mu - B_\mu{}^n\partial_n$ and the E$_{7(7)}$ connection coefficients
\begin{gather}
 \cQ_{\mu}^{A}{}_{B} = - \textstyle{\frac{1}{2}} \left[ {e^m}_a \partial_m B_{\mu}{}^{n} e_{n b} - ({e^p}_{a} \cD_{\mu} e_{p\, b}) \right] \Gamma^{ab}_{AB} 
 - \textstyle{\frac{\sqrt{2}}{12}} \Delta^{-1/2} {e'_{\mu}}{}^{\alpha} \left( F_{\alpha abc} \Gamma^{abc}_{AB} - \eta_{\alpha \beta \gamma \delta} F^{\beta \gamma \delta a} \Gamma_{a AB} \right), \\[3mm]
 \cP_{\mu ABCD} = \textstyle{\frac{3}{4}} \left[ {e^m}_a \partial_m B_{\mu}{}^{n} e_{n b} - ({e^p}_{a} \cD_{\mu} e_{p\, b}) \right] \Gamma^{a}_{[AB} \Gamma^{b}_{CD]}
 - \textstyle{\frac{\sqrt{2}}{8}} \Delta^{-1/2} {e'_{\mu}}{}^{\alpha} F_{abc \alpha} \Gamma^{a}_{[AB} \Gamma^{bc}_{CD]} \notag \\[2mm]
 \hspace{90mm} - \textstyle{\frac{\sqrt{2}}{48}} \Delta^{-1/2} e'_{\mu \, \alpha} \eta^{\alpha \beta \gamma \delta} F_{a \beta \gamma \delta}{\Gamma_{b}}_{[AB} \Gamma^{ab}_{CD]}.
\end{gather}
In analogy with this, the derivative of the new generalised vielbeine 
along the space-time directions satisfy 
\begin{align}\label{cDB}
&\cD_\mu e_{mnAB} + \frac{1}{2} \partial_{p} B_{\mu}{}^{p} e_{mnAB} + 2 \partial_{[m} 
B_{|\mu|}{}^{p} e_{n]pAB} + 3 \partial_{[m} B_{|\mu|np]} e^{p}_{AB} \notag \\
&\hspace{80mm}+ \cQ_{\mu}^{C}{}_{[A} e_{mnB]C} + \cP_{\mu ABCD} e_{mn}{}^{CD} = 0, \\[20pt]
&\cD_\mu e_{m_1 \dots m_5 AB} + \frac{1}{2} \partial_{p} B_{\mu}{}^{p} e_{m_1 \dots m_5 AB} - 5 \partial_{[m_1} B_{|\mu|}{}^{p} e_{m_2 \dots m_5]pAB} + 
\frac{3}{\sqrt{2}} \partial_{[m_1} B_{|\mu|m_2 m_3} e_{m_4 m_5]AB} \notag \\ \label{cDB2}
&- 6 \partial_{[m_1} \left (B_{|\mu| m_2 \dots m_5 p]} - \frac{\sqrt{2}}{4} B_{|\mu| m_2 m_3} 
A_{m_4 m_5 p]} \right) e^{p}_{AB} + \cQ_{\mu}^{C}{}_{[A} e_{m_1 \dots m_5 B]C} + \cP_{\mu ABCD} e_{m_1 \dots m_5}{}^{CD}=0, \\[10pt]
& \Bigg\{ \cD_\mu e_{m_1 \dots m_7, n AB} + \frac{1}{2} \partial_{p} B_{\mu}{}^{p} e_{m_1 \dots m_7, n AB} - 7 \partial_{m_1} B_{\mu}{}^{p} e_{ p m_2 \dots m_7, n AB} - \partial_{n} B_{\mu}{}^{p} e_{m_1 \dots m_7, p AB} \notag \\[2mm]
& \quad+ 3 \partial_{[n} B_{|\mu| m_1 m_2]} e_{m_3 \dots m_7AB} - 6 \partial_{[n} \left (B_{|\mu| m_1 \dots m_5]} - \frac{\sqrt{2}}{4} B_{|\mu| m_1  m_2} A_{m_3 m_4 m_5]} \right) e_{m_6 m_7 AB} \notag \\[2mm] \label{cDB3}
& \quad + \cQ_{\mu}^{C}{}_{[A} e_{m_1 \dots m_7, n B]C} + \cP_{\mu ABCD} e_{m_1 \dots m_7, n}{}^{CD} \; \Bigg\}_{[m_1 \dots m_7]} =0.
\end{align}
Note in particular that the vector fields that enter the 4-dimensional generalised vielbein postulate are precisely the E$_{7(7)}$ covariant vector fields \eqref{calB} that give rise 
to the generalised vielbeine.

In future work we intend to come back to relations \eqref{cDB}--\eqref{cDB3} and further investigate their role with regard to the embedding tensor formalism 
\cite{NSmaximal3, NScomgauge3} and the $D=4$ gaugings
studied in \cite{dWSTlag, dWSTmax4}.  As in the above
relations, where we are dealing with a ${\bf 56}$ of vector fields, there as well
the gauged theory is formulated in terms of a doubled set of 56 vector fields, such
that the 28 physical components are selected, together with the non-abelian
gauge group, by the embedding tensor, whose $D=11$ origins are expected to be hidden
in the above relations. This is of particular interest in view of the recent work
on the vacuum structure of maximal gauged supergravities in four dimensions, which
has turned out to be far richer than originally expected \cite{Fischbacher,DI} (see \cite{CDIZ} and 
references therein for more recent work on this).

\section{Outlook: generalisation to E$_{8(8)}$ and E$_{6(6)}$} \label{e8e6}

The results of this paper clearly point to an underlying structure of which
we have so far only seen a small part. In fact, similar results exist for other 
reductions of $D=11$ supergravity, most notably the one corresponding
to the 3+8 decomposition of the theory, where the relevant group is E$_{8(8)}$
and where, finally, the dual gravity field enters with full force, giving rise to eight
physical scalar degrees of freedom. For this case partial results have been known 
for a long time \cite{Nso16,KNS}. 

In this section  we briefly sketch 
how our construction generalises to E$_{8(8)}$  and also the simpler case 
of E$_{6(6)}$. In the former case, some of the relevant vielbeine have 
already been identified in \cite{Nso16,KNS}, and the existence of a corresponding 
E$_{8(8)}$-valued 248-bein in eleven dimensions is proved in \cite{KNS}, 
although in a more indirect manner. So let us consider this case first. To this aim, we perform a 
$3+8$ split of $D=11$. More specifically, the fields of the theory that give rise to 
scalar and vector degrees of freedom in a conventional reduction to three dimensions 
are, respectively, $g_{mn}$ and $A_{mnp}$, and  $B_{\mu}{}^{m}$ and $B_{\mu mn}$, 
where now $\mu= 0,1,2$ is a 3-dimensional space-time index and $m,n,p = 3, \dots, 10$ are 
the 8-dimensional spatial indices. As before, the field  $B_{\mu}{}^{m}$ is the off-diagonal 
component of the elfbein in the $3+8$ split, while $g_{mn}$ is the metric in 8-dimensional directions. $B_{\mu mn}$ is related to the eleven-dimensional 3-form by the field redefinition
$B_{\mu mn} = A_{\mu mn} - B_{\mu}{}^{p} A_{mnp}$, as before. As is well known,
in three dimensions vector fields are dual to scalars, so these fields account for 
the $248 - 120=128$ scalars that parametrise the E$_{8(8)}/$SU$(8)$ coset. 
It is shown in \cite{KNS} that the supersymmetry transformations of $B_{\mu}{}^{m}$ 
and $B_{\mu mn}$ in the SO(16) reformulation of $D=11$ supergravity \cite{Nso16} 
lead to two generalised vielbeine 
$$ e^{m}{}_{\mathcal{A}} \quad \textup{and} \quad e_{mn \mathcal{A}},$$
where $\mathcal{A}= 1, \dots , 248$ is an E$_{8(8)}$ index. In analogy with \eqref{cV1}
these generalised vielbeine can be combined into a $36 \times 248$ matrix, which
can be thought of as being part of a $248 \times 248$ E$_{8(8)}$ matrix. In fact, the 
existence of such an E$_{8(8)}$ matrix in eleven dimensions was inferred in \cite{KNS} 
by indirect group theoretic arguments. The results of this paper can now be used
to give a more explicit description of this matrix.

We will describe this construction elsewhere, but let us nevertheless outline the calculation
that needs to be done. In order to enlarge the $36 \times 248$ matrix we must consider 
a component  of the eleven-dimensional 6-form $B_{\mu m_1 \dots m_5} \sim B_\mu^{npq}.$ 
The supersymmetry  variation of this field leads to 56 further components, which 
add another $56\times 248$ chunk to the generalised vielbein. Finally, the dual gravity field
will give 64 further components from $B_{\mu m_1\dots m_7,n} \sim B_\mu{}^m{}_n$, 
which in total give a $156 \times 248$ matrix containing scalars coming from the reduction 
of the metric, 3-form, 6-form and dual gravity. Since these account for all of the scalar degrees 
of freedom, the completion of this matrix to an E$_{8(8)}$ matrix will not introduce any 
new degrees of freedom.~\footnote{See for example \cite{GGPE8}, where the 
E$_{8(8)}$ matrix is found by group theoretic means.} In other words, the full E$_{8(8)}$ 
matrix is completely determined by this 156 $\times$ 248 submatrix. However, there remains the interesting question of where the extra components come from. The GL(8,$\mathbb{R}$) decomposition
\begin{equation}
 {\bf 248} \; \longrightarrow \;  
 \bf{ \overline{8} + 28 + \overline{56} + 64 + 56 + \overline{28} + 8}
\end{equation}
suggests that three more fields are required in eleven dimensions in order to give rise to the remaining $56+28+8$ vector fields in the dimensionally reduced theory.

In addition, one can consider the status of the generalised vielbein postulate in this case.  Decomposing~\footnote{Our apologies to the reader for the multiple 
  different uses of these letters.}
\begin{equation*}
 \mathcal{A} = ([IJ],A),
\end{equation*}
where $I, J=1,\ldots,16$ and $A,B,...=1, \ldots, 128$ are now SO$(16)$ vector and chiral spinor indices, respectively, the generalised vielbein 
$e^{m}{}_{\mathcal{A}}$ satisfies \cite{Nso16}
\begin{gather}
\cD_\mu e^{m}_{IJ} + \partial_{n} B_{\mu}{}^{n} e^{m}_{IJ} + \partial_{n} B_{\mu}{}^{m} e^{n}_{IJ} +2 \cQ_{\mu K[I} e^{m}_{J]K} + \Gamma^{IJ}_{AB} \cP_{\mu}{}^{A} e^{m}_{B} = 0, \\[3mm]
\partial_{m} e^{n}_{IJ} +2 \cQ_{m K[I} e^{n}_{J]K} + \Gamma^{IJ}_{AB} \cP_{m}{}^{A} e^{n}_{B} = 0,
\end{gather}
where $\Gamma^{I}_{A\dot A}$ is a Spin$(16)$ gamma-matrix and the E$_{8(8)}$ connection components are defined in \cite{Nso16}.  The remaining components $e^{m}_{A}$ satisfy similar relations \cite{Nso16}.  Analogously, $e_{mn \mathcal{A}}$ found in \cite{KNS} satisfies
\begin{gather} \label{e83gvp2}
\cD_\mu e_{mnIJ} + \partial_{p} B_{\mu}{}^{p} e_{mnIJ} + 2 \partial_{[m} B_{\mu}{}^{p} e_{n]p IJ} + 18 \sqrt{2} \partial_{[m}B_{np]\mu} e^{p}_{IJ} + 2 \cQ_{\mu K[I} e_{mn J]K} + \Gamma^{IJ}_{AB} \cP_{\mu}{}^{A} e_{mn B} = 0, \\[3mm]
\partial_{p} e_{mn A} + 6 \sqrt{2} \left( \partial_p A_{mnq} - \frac{1}{4!} F_{pmnq} \right) e^{q}_{A} + \frac{1}{4} \cQ_{p IJ} \Gamma^{IJ}_{AB} e_{mn B} - \frac{1}{2} \Gamma^{IJ}_{AB} \cP_{p}{}^{B} e_{mn IJ} = 0. \label{e88gvp2}
\end{gather}
Note the striking resemblance of these equations to their E$_{7(7)}$ counterparts, equations \eqref{cDB} and \eqref{7gvp2}.  In particular, note the presence of the vector field $B_{\mu mn}$, the supersymmetry transformation of which gives $e_{mn \mathcal{A}}$ in equation \eqref{e83gvp2} and the non-gauge invariant ``connection'' term, analogous to connection \eqref{Xi1}, in equation \eqref{e88gvp2}.

The construction of the E$_{6(6)}$ matrix from the eleven-dimensional fields is more 
straightforward and only requires consideration of the eleven-dimensional metric, 3-form field
and its 6-form dual, because the dual gravity field does not give rise to any physical
degrees of freedom. In the $5+6$ split, the components of the eleven-dimensional fields that give rise to 
vector and scalar degrees of freedom under reduction to five dimensions are 
\begin{equation}
  B_{\mu}{}^{m}\,,\; B_{\mu mn}\,,\; B_{\mu \nu m}\,,\; g_{mn}\,,\; A_{mnp}\,,\; B_{\mu \nu \rho}, \label{e6fields}
\end{equation}
where now $\mu, \nu, \rho = 0, \dots 4$ and $ m,n,p= 5, \dots, 10.$ Note that in 5-dimensions, 3-forms are dual to scalars. Therefore, in total there are 42 scalars coming from $g_{mn}, A_{mnp}$ and $B_{\mu \nu \rho}$ that parametrise the E$_6/$USp$(8)$ coset. 

The E$_{6(6)}$ matrix in eleven dimensions can be constructed from the generalised vielbeine that arise from the supersymmetry transformations of $B_{\mu}{}^{m}, B_{\mu mn}, B_{\mu m_1 \dots m_5}$ in a USp(8) invariant reformulation of $D=11$ supergravity along the lines of \cite{dWNsu8, Nso16}. The E$_{6(6)}$ matrix is parametrised by $g_{mn}, A_{mnp}$ and the dual 6-form $A_{m_1 \dots m_6}.$ We stress once more that the construction of the E$_{6(6)}$ does not 
involve the dual gravity field and only depends on fields that are well-understood in eleven dimensions. The E$_{6(6)}$ matrix thus constructed should be equivalent to the 
E$_{6(6)}$ matrix constructed in \cite{BGPW} by group theory.

\vspace{1cm}\noindent
{\bf Acknowledgments:} We are grateful to Axel Kleinschmidt and Bernard de Wit for 
stimulating discussions, and to Bernard Schutz for inspiration \cite{schutz}.

\newpage
\appendix

\section{Conventions}

We use the following conventions:
$$ A_{[a_1 \dots a_{p}]} = \frac{1}{p!} \left( A_{a_1\dots a_p} + (p! - 1) \textup{ terms}\right),$$
$$(d A)_{a_1 \dots a_{p+1}} = (p+1)! \partial_{[a_1} A_{a_2 \dots a_{p+1}]},$$
$$ (\star A)_{a_1 \dots a_{d-p}} = \frac{i}{p!} \epsilon_{a_{1} \dots a_{d-p} b_{1} \dots b_{p} } A^{b_1 \dots b_p}.$$

\section{Supersymmetry transformation identities} \label{app:susytrans}
Below we list some equations that prove to be useful in deriving the supersymmetry transformations of the generalised vielbeine.   
\begin{align}
 \delta \left( i \Delta^{-1/2} \Gamma^p_{AB} \right) = -\sqrt{2} \, \Sigma_{ABCD} \left( i \Delta^{-1/2} \Gamma^p_{CD} \right)
 - 2 {\Lambda^C}_{[A} \left( i \Delta^{-1/2} \Gamma^p_{B]C} \right),
\end{align}
\begin{align}
 \delta \left( i \Delta^{-1/2} \Gamma_{mnAB} \right) =& -\sqrt{2} \, \Sigma_{ABCD} \left(- i \Delta^{-1/2} \Gamma_{mnCD} \right)
 - 2 {\Lambda^C}_{[A} \left( i \Delta^{-1/2} \Gamma_{mnB]C} \right) \notag \\[4mm] 
 &+\textstyle{\frac{3}{2}} \, i \Delta^{-1/2} \ \bar{\varepsilon} \Gamma_{[mn} \Psi_{p]} \; \Gamma^{p}_{AB},
\end{align}
\begin{align}
 \delta \left( i \Delta^{-1/2} \Gamma_{m_1 \ldots m_5 AB} \right) =& -\sqrt{2} \, \Sigma_{ABCD} \left( i \Delta^{-1/2} \Gamma_{m_1 \ldots m_5 CD} \right)
 - 2 {\Lambda^C}_{[A} \left( i \Delta^{-1/2} \Gamma_{m_1 \ldots m_5 B]C} \right) \notag \\[4mm] 
 &+15 \, i \Delta^{-1/2} \ \bar{\varepsilon} \Gamma_{[m_1 m_2} \Psi_{m_3} \Gamma_{m_4 m_5]AB} 
 +3 \, i \Delta^{-1/2} \ \bar{\varepsilon} \gamma_5 \Gamma_{[m_1 \ldots m_5} \Psi_{p]} \; \Gamma^{p}_{AB},
\end{align}
\begin{align}
 \delta \left(\Delta^{-1/2} \epsilon_{m_1 \ldots m_7}  \Gamma_{n AB} \right) =& -\sqrt{2} \, \Sigma_{ABCD} \left(- \Delta^{-1/2} \epsilon_{m_1 \ldots m_7}  \Gamma_{n CD} \right)
 - 2 {\Lambda^C}_{[A} \left(  \Delta^{-1/2} \epsilon_{m_1 \ldots m_7} \Gamma_{n B]C} \right) \notag \\[4mm] 
 &- \Delta^{-1/2} \epsilon_{m_1 \ldots m_7} \left(\textstyle{\frac{3}{4}} \bar{\varepsilon} \Gamma_{[pq} \Psi_{n]} \Gamma^{pq}_{AB} 
 +\textstyle{\frac{1}{2}} \bar{\varepsilon} \gamma_5 \Gamma^{pq} \Psi_{p} \; \Gamma_{q n AB}\right).
 \end{align}

The first equation in the list above is the precisely the supersymmetry transformation of the generalised vielbein $e^m_{AB}$ found in \cite{dWNsu8}. The second equation \cite{dWN13} is used to derive the supersymmetry transformation of the generalised vielbein $e_{mnAB}$.

\section{Six-form potential of the Englert solution} \label{app:englert}

In this appendix, we demonstrate that even for a very simple eleven-dimensional 
solution with non-vanishing flux, the Englert solution \cite{englert}, the six-form potential 
is non-zero and contains non-vanishing components mixing space-time and internal
components (thus vitiating one of the basic assumptions made in several recent
approaches to generalised geometry). This leads us to expect that the six-form potential 
will in general acquire a non-trivial form for all solutions with non-vanishing flux,
that is, all solutions other than the torus compactification.

The Englert solution satisfies the Freund-Rubin ansatz \cite{freundrubin} and preserves 
an SO$(7)^-$ subgroup of SO$(8)$.  More explicitly, the solution is of the form
\begin{eqnarray}
 g_{MN} &=&\gamma^{7/18}\left(\etao_{\mu \nu},\ \gamma^{-1/2} \go_{mn}\right), \quad
 \nonumber\\[1mm] 
F_{MNPQ}&=&\left(2 \sqrt{2} i \, m_7 \, \gamma^{5/6} \etao_{\mu \nu \rho \sigma},\ \frac{\sqrt{2}}{6} m_7 \, \gamma^{-1/6} \, \etao_{mnpqrst} \So^{rst}\right),
\end{eqnarray}
where $\etao_{\mu \nu}$ is the anti-de Sitter metric, $\go_{mn}$ is the round metric on the seven-sphere with inverse radius $m_7$ and all quantities with four-dimensional (seven-dimensional) indices are tensors with respect to $\etao_{\mu \nu}$ ($\go_{mn}$). $\gamma$ is an
arbitrary positive constant, which takes the value $ \gamma^{1/3}= 5/4$ when the solution is constructed
via the non-linear flux ansatz \cite{GGN}.
Furthermore, the torsion tensor $\So_{mnp}$ satisfies the relation
\begin{equation}
 \Do_{m}S_{npq}=\frac{1}{6} m_7 \etao_{mnpqrst} \So^{rst}.
\end{equation}

From equation \eqref{F7}, the six-form potential is given by the following equation
\begin{align} \label{def:A6englert}
7! D_{[M_1} A_{M_2 \dots M_7]} = \frac{i}{4!} \eta_{M_1 \ldots M_{11}} F^{M_8 \ldots M_{11}}  -  \frac{7!\sqrt{2}}{4! \, 2}  A_{[M_1 \dots M_3} F_{M_4 \dots M_7]}.
\end{align}
Note that
\begin{equation}
 \eta_{\mu \nu \rho \sigma m_1 \ldots m_7} = \gamma^{7/18} \etao_{\mu \nu \rho \sigma} \etao_{m_1 \ldots m_7},
\end{equation}
while
\begin{equation}
 F^{MNPQ} = \left( 2 \sqrt{2} i \, m_7 \, \gamma^{-13/18} \etao^{\mu \nu \rho \sigma},\ \frac{\sqrt{2}}{6} m_7 \, \gamma^{5/18} \, \etao^{mnpqrst} \So_{rst} \right),
\end{equation}
where the indices on $F_{MNPQ}$ are raised using the eleven-dimensional (inverse)
metric $g^{MN}$.  Clearly, the right hand side of equation \eqref{def:A6englert} is only non-zero for $[M_1 \ldots M_7]$ equal to $[m_1 \ldots m_7]$, $[\mu \nu \rho \sigma m n p]$ or $[\mu \nu \rho m_1 \ldots m_4]$.
Thus,
\begin{equation}
 7! \ D_{[M_1} A_{M_2 \dots M_7]} = \begin{cases}
				-\frac{15\sqrt{2}}{4} m_7 \, \gamma^{-1/3} \etao_{m_1 \ldots m_7} & [m_1 \ldots m_7] \\[1mm]
				\frac{\sqrt{2}}{2} i \, m_7 \, \gamma^{2/3} \etao_{\mu \nu \rho \sigma} \So_{mnp} & [\mu \nu \rho \sigma m n p] \\[1mm]
				- 2 \sqrt{2} i \, m_7 \, \gamma^{2/3}	\zetao_{\mu \nu \rho} \etao_{m_1 \ldots m_7} \So^{m_5 \ldots m_7} & [\mu \nu \rho m_1 \ldots m_4] \\[1mm]
				0	& \text{otherwise}
                                  \end{cases},
\end{equation}
where $\zetao_{\mu \nu \rho}$ is the potential for the Freund-Rubin field strength and is only defined locally
\begin{equation}
 4! \, \partial_{[\mu} \zetao_{\nu \rho \sigma]} = m_7 \, \etao_{\mu \nu \rho \sigma}.
\end{equation}
Hence,
\begin{equation}
 A_{M_1 \ldots M_6} = \begin{cases}
                       \frac{\sqrt{2}}{12}i \, \gamma^{2/3} \zetao_{\mu \nu \rho} \So_{mnp} & 
                      \quad  [\mu \nu \rho mnp] \\[1mm]
                     -\frac{15\sqrt{2}}{4} \gamma^{-1/3} \zetao_{m_1 \ldots m_6} &
                     \quad  [m_1 \ldots m_6] \\[1mm]
			0 & \quad  \text{otherwise}
                      \end{cases},
\end{equation}
where $\zetao_{m_1 \ldots m_6}$ is such that
\begin{equation}
 7! \, \partial_{[m_1} \zetao_{m_2 \ldots m_7]} = m_7 \, \etao_{m_1 \ldots m_7}.
\end{equation}
As anticipated, $A_{M_1\dots M_6}$ has non-vanishing components with {\em both}
space-time and internal indices.

\nocite{schutz}

\newpage

\bibliography{gv}

\providecommand{\href}[2]{#2}\begingroup\raggedright\begin{thebibliography}{10}

\bibitem{CJS}
E.~Cremmer, B.~Julia, and J.~Scherk, ``{Supergravity theory in
  eleven-dimensions},''
  \href{http://dx.doi.org/10.1016/0370-2693(78)90894-8}{{\em Phys.Lett.} {\bf
  B76} (1978)  409--412}.

\bibitem{dWN13}
B.~de~Wit and H.~Nicolai, ``{Deformations of gauged SO(8) supergravity and
  supergravity in eleven dimensions},''
  \href{http://dx.doi.org/10.1007/JHEP05(2013)077}{{\em JHEP} {\bf 1305} (2013)
   077},
\href{http://arxiv.org/abs/1302.6219}{{\tt arXiv:1302.6219 [hep-th]}}.

\bibitem{DIT}
G.~Dall'Agata, G.~Inverso, and M.~Trigiante, ``{Evidence for a family of SO(8)
  gauged supergravity theories},''
  \href{http://dx.doi.org/10.1103/PhysRevLett.109.201301}{{\em Phys.Rev.Lett.}
  {\bf 109} (2012)  201301},
\href{http://arxiv.org/abs/1209.0760}{{\tt arXiv:1209.0760 [hep-th]}}.

\bibitem{dWNsu8}
B.~de~Wit and H.~Nicolai, ``{d} = 11 supergravity with local {SU}(8)
  invariance,'' \href{http://dx.doi.org/10.1016/0550-3213(86)90290-7}{{\em
  Nucl.Phys.} {\bf B274} (1986)  363}.

\bibitem{cremmerjulia}
E.~Cremmer and B.~Julia, ``{The N=8 supergravity theory. 1. The Lagrangian},''
  \href{http://dx.doi.org/10.1016/0370-2693(78)90303-9}{{\em Phys.Lett.} {\bf
  B80} (1978)  48}.

\bibitem{Nso16}
H.~Nicolai, ``{D} = 11 supergravity with local {SO(16)} invariance,''
\href{http://dx.doi.org/10.1016/0370-2693(87)91102-6}{{\em Phys.Lett.} {\bf
  B187} (1987)  316}.

\bibitem{KNS}
K.~Koepsell, H.~Nicolai, and H.~Samtleben, ``{An exceptional geometry for D =
  11 supergravity?},''
  \href{http://dx.doi.org/10.1088/0264-9381/17/18/308}{{\em Class.Quant.Grav.}
  {\bf 17} (2000)  3689--3702}, \href{http://arxiv.org/abs/hep-th/0006034}{{\tt
  arXiv:hep-th/0006034 [hep-th]}}.

\bibitem{dWNN8}
B.~de~Wit and H.~Nicolai, ``{N=8 Supergravity},''
\href{http://dx.doi.org/10.1016/0550-3213(82)90120-1}{{\em Nucl.Phys.} {\bf
  B208} (1982)  323}.

\bibitem{dWNconsis}
B.~de~Wit and H.~Nicolai, ``{The consistency of the S**7 truncation in D=11
  supergravity},''
\href{http://dx.doi.org/10.1016/0550-3213(87)90253-7}{{\em Nucl.Phys.} {\bf
  B281} (1987)  211}.

\bibitem{NP}
H.~Nicolai and K.~Pilch, ``{Consistent truncation of d = 11 supergravity on
  AdS$_4 \times S^7$},'' \href{http://dx.doi.org/10.1007/JHEP03(2012)099}{{\em
  JHEP} {\bf 1203} (2012)  099},
\href{http://arxiv.org/abs/1112.6131}{{\tt arXiv:1112.6131 [hep-th]}}.

\bibitem{GGN}
H.~Godazgar, M.~Godazgar, and H.~Nicolai, ``{Testing the non-linear flux ansatz
  for maximal supergravity},''
  \href{http://dx.doi.org/10.1103/PhysRevD.87.085038}{{\em Phys.Rev.} {\bf D87}
  (2013)  085038},
\href{http://arxiv.org/abs/1303.1013}{{\tt arXiv:1303.1013 [hep-th]}}.

\bibitem{Hitchin}
N.~Hitchin, ``{Generalized Calabi-Yau manifolds},''
  \href{http://dx.doi.org/10.1093/qjmath/54.3.281}{{\em Quart.J.Math.Oxford
  Ser.} {\bf 54} (2003)  281--308},
\href{http://arxiv.org/abs/math/0209099}{{\tt arXiv:math/0209099 [math-dg]}}.

\bibitem{Gualtieri}
M.~Gualtieri, ``{Generalized complex geometry},''
\href{http://arxiv.org/abs/math/0401221}{{\tt arXiv:math/0401221 [math-dg]}}.

\bibitem{hullgenm}
C.~Hull, ``{Generalised Geometry for M-Theory},''
  \href{http://dx.doi.org/10.1088/1126-6708/2007/07/079}{{\em JHEP} {\bf 0707}
  (2007)  079},
\href{http://arxiv.org/abs/hep-th/0701203}{{\tt arXiv:hep-th/0701203
  [hep-th]}}.

\bibitem{pachwald}
P.~P. Pacheco and D.~Waldram, ``{M-theory, exceptional generalised geometry and
  superpotentials},''
  \href{http://dx.doi.org/10.1088/1126-6708/2008/09/123}{{\em JHEP} {\bf 0809}
  (2008)  123},
\href{http://arxiv.org/abs/0804.1362}{{\tt arXiv:0804.1362 [hep-th]}}.

\bibitem{westsl32}
P.~C. West, ``{E(11), SL(32) and central charges},''
  \href{http://dx.doi.org/10.1016/j.physletb.2003.09.059}{{\em Phys.Lett.} {\bf
  B575} (2003)  333--342},
\href{http://arxiv.org/abs/hep-th/0307098}{{\tt arXiv:hep-th/0307098
  [hep-th]}}.

\bibitem{hillmann}
C.~Hillmann, ``{Generalized E(7(7)) coset dynamics and D=11 supergravity},''
  \href{http://dx.doi.org/10.1088/1126-6708/2009/03/135}{{\em JHEP} {\bf 0903}
  (2009)  135}, \href{http://arxiv.org/abs/0901.1581}{{\tt arXiv:0901.1581
  [hep-th]}}.

\bibitem{BP}
D.~S. Berman and M.~J. Perry, ``{Generalized Geometry and M theory},''
  \href{http://dx.doi.org/10.1007/JHEP06(2011)074}{{\em JHEP} {\bf 1106} (2011)
   074},
\href{http://arxiv.org/abs/1008.1763}{{\tt arXiv:1008.1763 [hep-th]}}.

\bibitem{weste11}
P.~C. West, ``{E(11) and M theory},''
  \href{http://dx.doi.org/10.1088/0264-9381/18/21/305}{{\em Class.Quant.Grav.}
  {\bf 18} (2001)  4443--4460}, \href{http://arxiv.org/abs/hep-th/0104081}{{\tt
  arXiv:hep-th/0104081 [hep-th]}}.

\bibitem{BO}
A.~Borisov and V.~Ogievetsky, ``{Theory of Dynamical Affine and Conformal
  Symmetries as Gravity Theory},''
\href{http://dx.doi.org/10.1007/BF01038096}{{\em Theor.Math.Phys.} {\bf 21}
  (1975)  1179}.

\bibitem{west2000}
P.~C. West, ``{Hidden superconformal symmetry in M theory},'' {\em JHEP} {\bf
  0008} (2000)  007,
\href{http://arxiv.org/abs/hep-th/0005270}{{\tt arXiv:hep-th/0005270
  [hep-th]}}.

\bibitem{locale11}
F.~Riccioni and P.~West, ``{Local E(11)},''
  \href{http://dx.doi.org/10.1088/1126-6708/2009/04/051}{{\em JHEP} {\bf 0904}
  (2009)  051},
\href{http://arxiv.org/abs/0902.4678}{{\tt arXiv:0902.4678 [hep-th]}}.

\bibitem{dWNcc}
B.~de~Wit and H.~Nicolai, ``{Hidden symmetries, central charges and all
  that},'' \href{http://dx.doi.org/10.1088/0264-9381/18/16/302}{{\em
  Class.Quant.Grav.} {\bf 18} (2001)  3095--3112},
\href{http://arxiv.org/abs/hep-th/0011239}{{\tt arXiv:hep-th/0011239
  [hep-th]}}.

\bibitem{dufflu}
M.~Duff and J.~Lu, ``Duality rotations in membrane theory,''
\href{http://dx.doi.org/10.1016/0550-3213(90)90565-U}{{\em Nucl.Phys.} {\bf
  B347} (1990)  394--419}.

\bibitem{dft1}
C.~Hull and B.~Zwiebach, ``{Double Field Theory},''
  \href{http://dx.doi.org/10.1088/1126-6708/2009/09/099}{{\em JHEP} {\bf 0909}
  (2009)  099},
\href{http://arxiv.org/abs/0904.4664}{{\tt arXiv:0904.4664 [hep-th]}}.

\bibitem{dft2}
C.~Hull and B.~Zwiebach, ``{The Gauge algebra of double field theory and
  Courant brackets},''
  \href{http://dx.doi.org/10.1088/1126-6708/2009/09/090}{{\em JHEP} {\bf 0909}
  (2009)  090},
\href{http://arxiv.org/abs/0908.1792}{{\tt arXiv:0908.1792 [hep-th]}}.

\bibitem{dftgenmet}
O.~Hohm, C.~Hull, and B.~Zwiebach, ``{Generalized metric formulation of double
  field theory},'' \href{http://dx.doi.org/10.1007/JHEP08(2010)008}{{\em JHEP}
  {\bf 1008} (2010)  008},
\href{http://arxiv.org/abs/1006.4823}{{\tt arXiv:1006.4823 [hep-th]}}.

\bibitem{BGP}
D.~S. Berman, H.~Godazgar, and M.~J. Perry, ``{SO(5,5) duality in M-theory and
  generalized geometry},''
  \href{http://dx.doi.org/10.1016/j.physletb.2011.04.046}{{\em Phys.Lett.} {\bf
  B700} (2011)  65--67},
\href{http://arxiv.org/abs/1103.5733}{{\tt arXiv:1103.5733 [hep-th]}}.

\bibitem{BGPW}
D.~S. Berman, H.~Godazgar, M.~J. Perry, and P.~West, ``{Duality Invariant
  Actions and Generalised Geometry},''
  \href{http://dx.doi.org/10.1007/JHEP02(2012)108}{{\em JHEP} {\bf 1202} (2012)
   108},
\href{http://arxiv.org/abs/1111.0459}{{\tt arXiv:1111.0459 [hep-th]}}.

\bibitem{GGPE8}
H.~Godazgar, M.~Godazgar, and M.~J. Perry, ``{E8 duality and dual gravity},''
  \href{http://dx.doi.org/10.1007/JHEP06(2013)044}{{\em JHEP} {\bf 1306} (2013)
   044},
\href{http://arxiv.org/abs/1303.2035}{{\tt arXiv:1303.2035 [hep-th]}}.

\bibitem{BGGP}
D.~S. Berman, H.~Godazgar, M.~Godazgar, and M.~J. Perry, ``{The Local
  symmetries of M-theory and their formulation in generalised geometry},''
  \href{http://dx.doi.org/10.1007/JHEP01(2012)012}{{\em JHEP} {\bf 1201} (2012)
   012},
\href{http://arxiv.org/abs/1110.3930}{{\tt arXiv:1110.3930 [hep-th]}}.

\bibitem{CSW}
A.~Coimbra, C.~Strickland-Constable, and D.~Waldram, ``{$E_{d(d)} \times
  \mathbb{R}^+$ Generalised Geometry, Connections and M theory},''
\href{http://arxiv.org/abs/1112.3989}{{\tt arXiv:1112.3989 [hep-th]}}.

\bibitem{BCKT}
D.~S. Berman, M.~Cederwall, A.~Kleinschmidt, and D.~C. Thompson, ``{The gauge
  structure of generalised diffeomorphisms},''
  \href{http://dx.doi.org/10.1007/JHEP01(2013)064}{{\em JHEP} {\bf 1301} (2013)
   064},
\href{http://arxiv.org/abs/1208.5884}{{\tt arXiv:1208.5884 [hep-th]}}.

\bibitem{CSW2}
A.~Coimbra, C.~Strickland-Constable, and D.~Waldram, ``{Supergravity as
  Generalised Geometry II: $E_{d(d)} \times \mathbb{R}^+$ and M theory},''
\href{http://arxiv.org/abs/1212.1586}{{\tt arXiv:1212.1586 [hep-th]}}.

\bibitem{NSmaximal3}
H.~Nicolai and H.~Samtleben, ``{Maximal gauged supergravity in
  three-dimensions},''
  \href{http://dx.doi.org/10.1103/PhysRevLett.86.1686}{{\em Phys.Rev.Lett.}
  {\bf 86} (2001)  1686--1689},
\href{http://arxiv.org/abs/hep-th/0010076}{{\tt arXiv:hep-th/0010076
  [hep-th]}}.

\bibitem{NScomgauge3}
H.~Nicolai and H.~Samtleben, ``{Compact and noncompact gauged maximal
  supergravities in three-dimensions},'' {\em JHEP} {\bf 0104} (2001)  022,
\href{http://arxiv.org/abs/hep-th/0103032}{{\tt arXiv:hep-th/0103032
  [hep-th]}}.

\bibitem{dWSTlag}
B.~de~Wit, H.~Samtleben, and M.~Trigiante, ``{On Lagrangians and gaugings of
  maximal supergravities},''
  \href{http://dx.doi.org/10.1016/S0550-3213(03)00059-2}{{\em Nucl.Phys.} {\bf
  B655} (2003)  93--126},
\href{http://arxiv.org/abs/hep-th/0212239}{{\tt arXiv:hep-th/0212239
  [hep-th]}}.

\bibitem{englert}
F.~Englert, ``Spontaneous compactification of eleven-dimensional
  supergravity,''
\href{http://dx.doi.org/10.1016/0370-2693(82)90684-0}{{\em Phys.Lett.} {\bf
  B119} (1982)  339}.

\bibitem{CJLP}
E.~Cremmer, B.~Julia, H.~Lu, and C.~Pope, ``{Dualization of dualities. 1.},''
  \href{http://dx.doi.org/10.1016/S0550-3213(98)00136-9}{{\em Nucl.Phys.} {\bf
  B523} (1998)  73--144},
\href{http://arxiv.org/abs/hep-th/9710119}{{\tt arXiv:hep-th/9710119
  [hep-th]}}.

\bibitem{hulldual}
C.~Hull, ``{Duality in gravity and higher spin gauge fields},'' {\em JHEP} {\bf
  0109} (2001)  027,
\href{http://arxiv.org/abs/hep-th/0107149}{{\tt arXiv:hep-th/0107149
  [hep-th]}}.

\bibitem{BBH}
X.~Bekaert, N.~Boulanger, and M.~Henneaux, ``{Consistent deformations of dual
  formulations of linearized gravity: a no go result},''
  \href{http://dx.doi.org/10.1103/PhysRevD.67.044010}{{\em Phys.Rev.} {\bf D67}
  (2003)  044010},
\href{http://arxiv.org/abs/hep-th/0210278}{{\tt arXiv:hep-th/0210278
  [hep-th]}}.

\bibitem{BdRKKR}
E.~A. Bergshoeff, M.~de~Roo, S.~F. Kerstan, A.~Kleinschmidt, and F.~Riccioni,
  ``{Dual gravity and matter},''
  \href{http://dx.doi.org/10.1007/s10714-008-0650-4}{{\em Gen.Rel.Grav.} {\bf
  41} (2009)  39--48},
\href{http://arxiv.org/abs/0803.1963}{{\tt arXiv:0803.1963 [hep-th]}}.

\bibitem{NicTownsvN}
H.~Nicolai, P.~Townsend, and P.~van Nieuwenhuizen, ``Comments on
  eleven-dimensional supergravity,''
{\em Lett.Nuovo Cim.} {\bf 30} (1981)  315.

\bibitem{DAuriaFre}
R.~D'Auria and P.~Fre, ``Geometric supergravity in d = 11 and its hidden
  supergroup,''
\href{http://dx.doi.org/10.1016/0550-3213(82)90376-5}{{\em Nucl.Phys.} {\bf
  B201} (1982)  101--140}.

\bibitem{TownsDemoc}
P.~Townsend, ``{P-brane democracy},''
\href{http://arxiv.org/abs/hep-th/9507048}{{\tt arXiv:hep-th/9507048
  [hep-th]}}.

\bibitem{deAlwis}
S.~de~Alwis, ``{Coupling of branes and normalization of effective actions in
  string / M theory},'' \href{http://dx.doi.org/10.1103/PhysRevD.56.7963}{{\em
  Phys.Rev.} {\bf D56} (1997)  7963--7977},
\href{http://arxiv.org/abs/hep-th/9705139}{{\tt arXiv:hep-th/9705139
  [hep-th]}}.

\bibitem{BBS}
I.~A. Bandos, N.~Berkovits, and D.~P. Sorokin, ``{Duality symmetric
  eleven-dimensional supergravity and its coupling to M-branes},''
  \href{http://dx.doi.org/10.1016/S0550-3213(98)00102-3}{{\em Nucl.Phys.} {\bf
  B522} (1998)  214--233},
\href{http://arxiv.org/abs/hep-th/9711055}{{\tt arXiv:hep-th/9711055
  [hep-th]}}.

\bibitem{curtright}
T.~Curtright, ``Generalized gauge fields,''
\href{http://dx.doi.org/10.1016/0370-2693(85)91235-3}{{\em Phys.Lett.} {\bf
  B165} (1985)  304}.

\bibitem{dualHull1}
C.~Hull, ``{Strongly coupled gravity and duality},''
  \href{http://dx.doi.org/10.1016/S0550-3213(00)00323-0}{{\em Nucl.Phys.} {\bf
  B583} (2000)  237--259},
\href{http://arxiv.org/abs/hep-th/0004195}{{\tt arXiv:hep-th/0004195
  [hep-th]}}.

\bibitem{Obers:1997kk}
N.~Obers, B.~Pioline, and E.~Rabinovici, ``{M theory and U duality on T**d with
  gauge backgrounds},''
  \href{http://dx.doi.org/10.1016/S0550-3213(98)00264-8}{{\em Nucl.Phys.} {\bf
  B525} (1998)  163--181},
\href{http://arxiv.org/abs/hep-th/9712084}{{\tt arXiv:hep-th/9712084
  [hep-th]}}.

\bibitem{Obers:1998fb}
N.~Obers and B.~Pioline, ``{U duality and M theory},''
  \href{http://dx.doi.org/10.1016/S0370-1573(99)00004-6}{{\em Phys.Rept.} {\bf
  318} (1999)  113--225},
\href{http://arxiv.org/abs/hep-th/9809039}{{\tt arXiv:hep-th/9809039
  [hep-th]}}.

\bibitem{DHN}
T.~Damour, M.~Henneaux, and H.~Nicolai, ``{E(10) and a 'small tension
  expansion' of M theory},''
  \href{http://dx.doi.org/10.1103/PhysRevLett.89.221601}{{\em Phys.Rev.Lett.}
  {\bf 89} (2002)  221601},
\href{http://arxiv.org/abs/hep-th/0207267}{{\tt arXiv:hep-th/0207267
  [hep-th]}}.

\bibitem{dWNW}
B.~de~Wit, H.~Nicolai, and N.~P. Warner, ``The embedding of gauged {N}=8
  supergravity into d = 11 supergravity,''
\href{http://dx.doi.org/10.1016/0550-3213(85)90128-2}{{\em Nucl.Phys.} {\bf
  B255} (1985)  29}.

\bibitem{dWSTmax4}
B.~de~Wit, H.~Samtleben, and M.~Trigiante, ``{The Maximal D=4
  supergravities},''
  \href{http://dx.doi.org/10.1088/1126-6708/2007/06/049}{{\em JHEP} {\bf 0706}
  (2007)  049},
\href{http://arxiv.org/abs/0705.2101}{{\tt arXiv:0705.2101 [hep-th]}}.

\bibitem{NFlevel}
H.~Nicolai and T.~Fischbacher, ``{Low level representations for E(10) and
  E(11)},''
\href{http://arxiv.org/abs/hep-th/0301017}{{\tt arXiv:hep-th/0301017
  [hep-th]}}.

\bibitem{Fischbacher}
T.~Fischbacher, ``The encyclopedic reference of critical points for
  {SO}(8)-gauged {N}=8 supergravity. part 1: Cosmological constants in the
  range $-\lambda / g^2 \in $[6:14.7),''
\href{http://arxiv.org/abs/1109.1424}{{\tt arXiv:1109.1424 [hep-th]}}.

\bibitem{DI}
G.~Dall'Agata and G.~Inverso, ``{On the Vacua of N = 8 Gauged Supergravity in 4
  Dimensions},'' \href{http://dx.doi.org/10.1016/j.nuclphysb.2012.01.023}{{\em
  Nucl.Phys.} {\bf B859} (2012)  70--95},
\href{http://arxiv.org/abs/1112.3345}{{\tt arXiv:1112.3345 [hep-th]}}.

\bibitem{CDIZ}
F.~Catino, G.~Dall'Agata, G.~Inverso, and F.~Zwirner, ``{On the moduli space of
  spontaneously broken N = 8 supergravity},''
\href{http://arxiv.org/abs/1307.4389}{{\tt arXiv:1307.4389 [hep-th]}}.

\bibitem{schutz}
B.~F. Schutz, {\em {Gravity from the ground up}}.
\newblock Cambridge University Press,
2003.
\newblock

\bibitem{freundrubin}
P.~G. Freund and M.~A. Rubin, ``Dynamics of dimensional reduction,''
\href{http://dx.doi.org/10.1016/0370-2693(80)90590-0}{{\em Phys.Lett.} {\bf
  B97} (1980)  233--235}.

\end{thebibliography}\endgroup
\bibliographystyle{utphys}
\end{document}